\documentclass{article}

\usepackage{arxiv}

\usepackage{amssymb}
\usepackage{graphics, graphicx}

\usepackage{siunitx}
\usepackage{array}
\usepackage{makecell, booktabs, multirow}

\usepackage{mwe}
\usepackage{amsmath,amssymb,amsfonts}
\usepackage{algorithmic}
\usepackage{graphicx}
\usepackage{textcomp}
\usepackage{xcolor}
\usepackage{soul}
\usepackage{caption}
\usepackage{subcaption}
\usepackage{tabularx,tikz}

\usepackage{stackengine}

\usepackage{amsthm}
\usepackage{pgfplots}
\pgfplotsset{compat = newest}
 
\theoremstyle{definition}
\newtheorem{definition}{Definition}[section]

\usepackage{authblk}
\usepackage{blindtext}

\title{A Novel Methodology for Designing Policies in \\Mobile Crowdsensing Systems}
\author[ \hspace{0.1cm}, 1, 2]{Alessandro~Di~Stefano
\thanks{Corresponding Author: distefano.alessandro@gmail.com}}
\author[3]{Marialisa~Scat\`{a}}
\author[3]{Barbara~Attanasio}
\author[3]{Aurelio~La~Corte}
\author[1]{Pietro~Li\`o}
\author[4]{Sajal~K.~Das}

\affil[1]{\textit{Computer Laboratory, Department of Computer Science, University of Cambridge, 15 JJ Thomson Avenue, Cambridge CB3 0FD, UK}}
\affil[2]{\textit{School of Computing, Engineering and Digital technologies (SCEDT), Department of Computing and Games, Teesside University, Middlesbrough TS1 3BX, UK}}
\affil[3]{\textit{Department of Electrical, Electronic and Computer Engineering (DIEEI), University of Catania, 6 Viale A. Doria, Catania, 95125, Italy}}
\affil[4]{\textit{Department of Computer Science, Missouri University of Science and Technology, 15 Computer Science Building, 500 W. 15th Street, MO 65409 Rolla, USA}}

\begin{document}

\maketitle

\begin{abstract}
Mobile crowdsensing is a people-centric sensing system based on users' contributions and incentive mechanisms aim at stimulating them. In our work, we have rethought the design of incentive mechanisms through a game-theoretic methodology. Thus, we have introduced a multi-layer social sensing framework, where humans as social sensors interact on multiple social layers and various services. We have proposed to weigh these dynamic interactions by including the concept of homophily, that is a human-related factor related to the similarity and frequency of interactions on the multiplex network. We have modelled the evolutionary dynamics of sensing behaviours by defining a mathematical framework based on multiplex EGT, quantifying the impact of homophily, network heterogeneity and various social dilemmas. We have detected the configurations of social dilemmas and network structures that lead to the emergence and sustainability of human cooperation. Moreover, we have defined and evaluated local and global Nash equilibrium points by including the concepts of homophily and heterogeneity. Therefore, we have analytically defined and measured novel statistical measures of social honesty, QoI and users' behavioural reputation scores based on the evolutionary dynamics.
Through the proposed methodology we have defined the Decision Support System (DSS) and a novel incentive mechanism by operating on the policies in terms of users' reputation scores, that also incorporate users' behaviours other than quality and quantity of contributions.
To evaluate our methodology experimentally, we consider a real dataset on vehicular traffic monitoring crowdsensing application, Waze, and we have derived the disbursement of incentives by also comparing our method with baselines. Experimental results demonstrate that our methodology, based on both quality and quantity of reports and the local or microscopic spatio-temporal distribution of behaviours, is able to better discriminate users' behaviours. This multi-scale characterisation of users (both global and local) represents a novel research direction and paves the way for novel policies on mobile crowdsensing systems.
\end{abstract}


\section{Introduction}
\label{sec:intro}
Recent years have witnessed a widespread diffusion and adoption of mobile devices (e.g., smartphones, tablets) in our daily lives \cite{sajal2020,chen2019designing,zhang2015incentives,boubiche2018mobile}. In addition, the integration of a wide variety of embedded sensors (e.g., GPS, gyroscope, camera, etc.) into mobile devices along with improved processing power and storage capacities, have enhanced their sensing capabilities \cite{chen2019designing,boubiche2018mobile}. In parallel, the growth of mobile networks and wireless communication technologies have resulted in a better connectivity between user devices and vehicular systems \cite{boubiche2018mobile}. Such enhanced mobile and pervasive technologies have led to a a wide range of applications \cite{sajal2020,boubiche2018mobile} that are part of the paradigm termed as {\em Mobile Crowdsensing} (MCS). Thus,
MCS is a people-centric sensing system based on human voluntary participation or contribution about some phenomena in their surrounding environment. For example, human users with their personal mobile devices acquire local information (e.g., geo-spatial) and share their knowledge or measures with other users and communities in the network \cite{sajal2020,zhang2015incentives,jin2018thanos}. This large-scale sensing paradigm leverages the collaborative approach and contributes to measure phenomena of mutual interest \cite{sajal2020,zhang2015incentives,yang2016incentive}.

MCS is also an excellent example of the cyber-physical convergence phenomenon, leading to the {\em Internet of People} (IoP) paradigm \cite{conti2017internet}.
Humans carrying mobile devices not only act as participatory or ``social sensors", gathering data, but they also interact with the physical and cyber worlds to accomplish changes \cite{mordacchini2017social,socialnetworking}. Thus, the dynamics of human behaviours play a key role in better understanding the complex behaviour of the cyber-physical-human world, putting people at the centre of this novel IoP paradigm.
However, since participating in the sensing systems may incur costs and risks, common individuals are unwilling to participate and feed the system with their sensed data due to the lack of sufficient incentives or pushes towards {\em cooperation}. Consuming computational and communication resources of the personal smart devices, or privacy-related issues concerned with the provided location information when collecting data, are only some of these risks/costs. It becomes therefore crucial to motivate users with incentive mechanisms \cite{sajal2020,zhang2015incentives,restuccia2016incentive,gao2015survey,restuccia2018incentme}, encouraging them to provide their sensing contributions in a timely and reliable manner. It is important to observe how both the number (i.e., quantity) and the accuracy (i.e., quality) of reports assume a key role in the operational reliability of a MCS application \cite{sajal2020}.

In this context, we propose a game-theoretic methodology in order to define a decision support system and the design of a novel incentive mechanism. Its definition is based on some statistical measures, that is the Quality of Information (QoI) and the reputation scores of each user in the network. These estimators are derived from the evolutionary dynamics of human sensing behaviours on a multi-layer social sensing framework, where we quantify the impact of homophily, network heterogeneity and multiplexity. Indeed, by exploring the evolutionary dynamics of human cooperation, we detect which configurations of social dilemmas and network structures lead to the emergence and resilience of cooperation in a complex network scenario. 
In the experimental validation, we consider a real-world dataset available from a popular but proprietary vehicular traffic management apps called \textit{Waze}. We derive the disbursement of incentives by also comparing our method with baselines. We find how our methodology is able to better differentiate users' behaviours based on both quality and quantity of reports (global) other than the local or microscopic spatio-temporal distribution of behaviours. Fig. \ref{fig:scheme} describes the framework and the various steps of the proposed modelling procedure.
\vspace{-4pt}
\subsection{Motivations}
To the best of our knowledge, most of existing approaches to derive incentive mechanisms used in mobile crowdsensing applications are based on stimulating the degree of participation of users with regular contributions. Hence, incentives are disbursed based on the ``quantity'' or degree of participation of users, without considering the ``quality'' of information, namely the accuracy or truthfulness of contributions \cite{sajal2020,barnwal2019publish,restuccia2017quality}. As underlined in \cite{sajal2020,barnwal2019publish}, we must take into account both aspects, since false contributions may result in publishing wrong information, dramatically influencing the operational reliability of the MCS service. 
Following \cite{sajal2020,barnwal2019publish}, we argue that besides the quantity, we need a measure that assesses the {\em Quality of Information (QoI)}, that is the trustworthiness of the aggregated contributions generated from users. User reputation scores must be based on this QoI measure and their definition and measure allow to (a) distinguish honest from selfish or malicious users, (b) disburse the right incentives to users, and (c) support decision making about publishing or dropping information in a MCS application scenario. Now, let us synthesise some of the main limitations of existing QoI, reputation models and incentive mechanisms, also in relation with existing socially-aware game-theoretic approaches.

First, QoI and reputation models are usually based on Beta and Dirichlet distributions (such as J\o sang's belief model and Dempster-Shafer (D-S) reputation), or their variants \cite{barnwal2019publish}. But both these models are not able to fairly capture the degree of participation (quantity) and quality together into the reputation score. As underlined in \cite{sajal2020}, sometimes they end up giving excessive importance to the degree of participation, neglecting the importance of quality, or vice versa.
To deal with this challenging issue, coherently with \cite{sajal2020,barnwal2019publish}, we derive a QoI score and a reputation score that include both quality and quantity of contributions. This step is crucial to weaken the success rates of malicious ratings and make the system more robust and reliable \cite{sajal2020}.

Second, the social aspects related to human sensing behaviours and the impact of interactions between users are becoming increasingly important in the evaluation of incentive mechanisms. Some proposed QoI-aware incentive mechanisms account for quality of each sensing report before assigning incentives \cite{sajal2020,restuccia2018incentme}. 
Furthermore, recently some game-theoretic (e.g. auction-based, Stackelberg games, etc.) and also socially-aware incentive mechanisms have been proposed aiming at analysing and increasing the participation level of users, also including social network effects and strategic participation of users in MCS applications \cite{zhang2015incentives,nie2019stackelberg,gong2015exploiting,nie2018socially}.
Nevertheless, the main limitations consist of the assumption of rationality of agents and neglecting the presence of selfish or malicious participants in the sensing task. In addition, their modelling approach is not inherently dynamic, and they include only one layer of analysis (single-layer network). More realistically, in a social network entities/human users interact simultaneously through multiple types of social interactions. In order to include the evolutionary dynamics and this more realistic description of social interactions, we have proposed a multiplex EGT-based model in our game-theoretic methodology, targeted at exploring human cooperation on a multiplex social sensing.
In addition, we analyse the role of the various social dilemmas and complex network topologies. Indeed, considering various complex network structures with different topologies allows us to also encompass the effect of structural heterogeneity in the evolution of cooperation. 
Finally, as explained in the model (see section \ref{sec:model}), we have also included a measure of homophily, to account for the role of the frequency of interactions between similar nodes in their sensing behaviours. Based on the concepts of homophily and heterogeneity we have defined and evaluated local and global Nash equilibrium points.

Third, both the design of incentive mechanisms and reducing incentive losses play a vital role in mobile crowdsensing applications. Based on the above reasoning, these mechanisms should be fair and dynamic.
These features are associated with adaptive rewards for users, that is incentives changing over time based on their contributions, that should be evaluated in terms of both quality and quantity. 
To include both fairness and dynamics in the design of incentive mechanisms, we have rethought them by introducing a novel game-theoretic methodology, based on modelling and quantifying the evolutionary dynamics of human sensing behaviours (see section \ref{sec:model}).
Our definitions of `social honesty', QoI, user reputation scores and incentives encompass both these features, where these statistical estimators are derived from the cooperativeness of each human user in the network.
This leads us to analytically redefine QoI and reputation scores, representing key aspects for defining the Decision Support System (DSS) on which the design of incentive mechanisms is based. Moreover, we have redefined the incentive mechanism, based not only on quality and quantity of reports, but we have also characterised users according to local or microscopic spatio-temporal distribution of social behaviours (see subsection \ref{ssec:DSS}). The final aim is to define an increasingly differentiated and inherently dynamic incentive mechanism (see subsection \ref{ssec:performance}).
\subsection{Contributions of this Paper}
\label{ssec:contributions}
Once explained the motivations behind the proposed methodology, the main contributions of the paper are summarised below.
\begin{itemize}
    \item We redesign the problem of incentive mechanism by introducing a novel game-theoretic methodology. It allows us to model and quantify the evolutionary dynamics of human sensing behaviours. We define a multi-layer social sensing framework to explore and quantify the dynamics patterns of interactions due to multiple layers, inter-layer coupling and communicability. We also quantify how the evolution of cooperation hinges on the number of layers. 
    \item We measure the joint impact of network properties and human-related factors, such as homophily, in the evolutionary dynamics. To this aim, we define weighted connections between nodes, where weights are defined by taking into account both the concept of homophily and centrality measure in the multiplex structure. We point out how these weights influence behaviours on the multi-layer social sensing framework. 
    \item To understand the role of network structural heterogeneity on the evolution of human cooperation, we explore and quantify how the various network structures with a different degree of heterogeneity impact on the evolutionary dynamics of users' sensing behaviours. We detect how and to what extent some specific network structures and games lead to the emergence and sustainability of cooperation in a complex network scenario.
    \item We analytically define novel statistical estimators related to trustworthiness, that is the Quality of Information (QoI) and the aggregated behavioural reputation scores of each user in the network. These statistical measures, together with quality and quantity of contributions, allow us to define an overall measure of composite user reputation scores, the Decision Support System (DSS) and design a novel incentive mechanism. 
    \item We validate our methodology using a real dataset from Waze and derive the disbursement of incentives based on our model. We shed light on how our model is able to better discriminate users by including a multi-scale (both global and local) characterisation of users. Indeed, along with quality and quantity of reports, we include the local or microscopic spatio-temporal distribution of behaviours.
\end{itemize}

The remainder of the paper is organised as follows. Section \ref{sec:related_work} summarises recent literature works on the various aspects considered in our work. 
Section \ref{sec:model} present the system model and deals with the novel social game-theoretic methodology proposed in this work. Section \ref{sec:results} presents the simulation analysis and analytical results.
Section \ref{sec:experimental_validation} describes the experimental validation on Waze dataset.
Finally, 
Section \ref{sec:discussions} discusses findings and some aspects of the proposed methodology, and Section \ref{sec:conclusions} concludes the paper with some directions of future research.
\section{Related Work}
\label{sec:related_work}
Recent literature works have shed light on the role of network heterogeneity, homophily, also providing some key insights on how and when cooperation emerges and sustain in the network considering the different social dilemmas.
In this section, we review the existing works, other than some proposed game-theoretic models used to design socially-aware incentive mechanisms in MCS. 
\subsection{Multiplexity, Homophily,  Heterogeneity and Social Dilemmas in the Evolution of Cooperation}
Relationships between individuals and users are typically multi-relational in nature so that social behaviours and their dynamics can be understood only by taking into account more than a single network of interactions between them. In recent years multi-layer networks \cite{di2015quantifying,battiston2017new,boccaletti14} have been introduced and regarded as the most suitable way to describe social networks or transportation networks, only to cite a few examples. 
In \textit{multiplex networks}, social interactions occur on different contexts and social environments (e.g., family, friendships, colleagues, etc.) and an individual’s behaviour can be different on each layer, although it is determined from the simultaneous interactions on all the layers of the multiplex structure \cite{di2015quantifying,battiston2017new,scata2018quantifying,boccaletti14,di2019improving}.
In an MCS system, the multiplex nature of social interactions of human users carrying their devices therefore adds a further level of complexity. 
Indeed, other than intra-layer relationships on each single network (or layer), we also need to encompass the inter-layer interactions between humans and their counterpart nodes on the other layers of the multiplex structure. Only by studying the inter-layer interactions between nodes, it is possible to detect the emergent behaviours and focus on the key features related to nodes and edges, from which these patterns are generated \cite{boccaletti14}.

\textit{Evolutionary Game Theory} (EGT) constitutes the most common framework adopted to face with the conundrum of human cooperation, and social dilemmas are typically used as general metaphors for studying the evolution of cooperation \cite{hofbauer2003evolutionary,di2015quantifying,gomez2012evolution}. \textit{Social dilemmas} describe all the conflict situations where the strategy with the highest individual fitness does not represent the most convenient strategy in terms of social community \cite{scata2016combining}. Thus, the players and the society would benefit more from mutual \textit{cooperation}, yielding both an individual and total benefit higher than that of mutual defection. Even though cooperation may not emerge under these conditions, in nature and real-world networks we can observe how the cooperation does exist. One of the main targets of our work is to better understand the underlying mechanisms in terms of network structure and human-related factors driving cooperation in a networked scenario.

{\em Homophily} is the principle for which similarity breeds connection, namely the tendency to associate and interact more frequently with similar people, as extensively explained in \cite{di2015quantifying,scata2018quantifying,scata2016combining}.
In terms of evolutionary dynamics of human cooperation, in \cite{di2015quantifying} authors have demonstrated how homophily plays a key role in shaping human behaviours, by guiding and speeding up the emergence of cooperation between individuals. In a \textit{multi-layer network}, homophily represents the degree correlation of two nodes on different layers, and it may become a key factor in guiding social behaviours. So far, the role of homophily in the evolution of cooperation on a social multiplex network has been explored and quantified considering only a Scale-Free (SF) network topology and a fixed number of layers \cite{di2015quantifying}. Given real-world populations are heterogeneous, there are individuals (hubs in a SF network) having many more connections than others. 

Some other literature works \cite{santos2006evolutionary,poncela2007robustness,dercole2019direct} have unveiled how network structural \textit{heterogeneity} plays a key role in the evolution of cooperation as it enhances the emergence and resilience of cooperation. They have demonstrated how the sustainability of cooperation is simpler to achieve in heterogeneous rather than in homogeneous populations, regardless the social dilemma considered as a metaphor for investigating human cooperation. 
In the field of multiplex networks there are a few results about the role of multiplex networks and structural heterogeneity on the \textit{evolution of cooperation} \cite{gomez2012evolution,matamalas2015strategical}. It is important to note that their results have been derived in most cases by using the mean-field hypothesis and assuming there is no correlation between the strategies used by an individual in each layer of the multiplex structure \cite{gomez2012evolution}. Instead, in our work, we include correlation between nodes’ strategies on different layers, which depends on the communicability function, capturing inter-layer coupling and \textit{interdependence} between layers \cite{di2015quantifying,estrada2014communicability}.
Moreover, since the layers of a multiplex structure may exhibit a different network topology and degree of heterogeneity, in this work we explore and quantify how the various \textit{network structures} with a different structural heterogeneity impact on the evolutionary dynamics of users' sensing behaviours.

In \cite{matamalas2015strategical} the authors have systematically studied the evolution of cooperation in four social dilemmas that we also consider in the T-S plane on the multiplex network.
They found out some features in the microscopic organisation of strategies, that are responsible for the important differences between cooperative dynamics in monoplex or single-layer networks and multiplex networks. Moreover, some works have demonstrated how the extent of multiplexity, which hinges also on the number of layers and inter-layer coupling measure between them affect the emergent social dynamics on the multiplex network \cite{matamalas2015strategical,gomez2012evolution,di2015quantifying}. Also, the role of homophily may result different according to the underlying network structure of each layer and the number of layers composing the multiplex structure. 
It is important to note that in their work authors have considered that each layer is a homogeneous graph (i.e., Erd\"os-R\'enyi (ER)) and they have adopted the replicator-like rule for the nodes’ strategy update. Differently from \cite{matamalas2015strategical}, we take into account various network structures (see \ref{ssec:dilemmas_topologies}), exhibiting a different structural heterogeneity, and the Fermi rule as microscopic strategy update rule (see \ref{ssec:evolutionary}). Some other authors \cite{iyer2016evolution} have analysed the evolutionary dynamics of various social dilemmas, deriving Nash equilibrium points before under the hypothesis of well-mixed population and then considering a ER-structured population. In the latter, they have distinguished between an interaction network and an updating network, thus separating the role of layers in the multiplex structure. In our work, we do not separate the role of each layer and, in addition, we measure the impact of network heterogeneity and homophily (see \ref{ssec:multiplex_social_sensing}).
From literature works, it is clear how there are some unsolved questions in the field of evolutionary dynamics of human behaviours on a social multiplex network. Indeed, in the hypothesis of considering a \textit{structured population}, where nodes interact with their neighbours on each layer of the multiplex network and on each layer of the \textit{multi-layer sensing framework}, we will answer to the following question: \textit{what is the joint impact of network heterogeneity, homophily and multiplexity on a structured population in terms of evolutionary dynamics of human sensing behaviours?}
\subsection{Game-theoretic and Socially-aware Incentive Mechanisms}
Most of the incentive mechanisms used in CS applications are aimed at stimulating the degree and regularity of contributions. Overall, the factor mainly used to decide how to disburse incentives is therefore related to ``quantity'' (i.e., degree of participation), without taking into account the ``quality'' of information, that is the accuracy or truthfulness of contributions \cite{sajal2020,restuccia2017quality}. As underlined in \cite{sajal2020}, we must take into account both aspects, since false contributions may result in publishing wrong information, dramatically influencing the service operation. 
Thus, it becomes essential to measure the {\em Quality of Information (QoI)} related to users' contributions and then derive a {\em behavioural user reputation score} of each human user in the MCS application.

Some recent works have proposed game-theoretic models and socially-aware incentive mechanisms in MCS scenarios in order to analyse and increase the participation level of users, also including social network effects \cite{zhang2015incentives,nie2019stackelberg,nie2018socially,gong2015exploiting}. However, none of them has ever included multiplex network and homophily and a game-theoretic model targeted at exploring human cooperation on a multiplex social sensing by analysing the role of the various social dilemmas and complex network topologies. 
Various approaches based on game-theoretic models have addressed the issue of strategic participation of users in a MCS application 
\cite{sajal2020,xiao2018mobile}. Recently, in \cite{nie2019stackelberg} authors have proposed to consider also the social structure information and influence between users in a socially-aware Bayesian Stackelberg game to explore the users' participation level. Then, they have introduced a backward induction to propose an optimal incentive mechanism of the crowdsensing service provider. Also in other works \cite{nie2018socially,gong2015exploiting} it has been investigated how to exploit the social network effects and trust to encourage users' participation through reciprocity, so that crowdsensing service providers obtain a greater gain. In most of the CS incentive schemes, auctions and pricing strategies provide incentives to mobile users to participate in crowdsensing applications. The basic idea is to either maximise the total utility/value of the sensing platform under certain costs/budgets constraints or minimise the total disbursement of the platform. 

Nevertheless, some of the main drawbacks of the proposed game-theoretic models are mostly related to the lack of a dynamic modelling approach, the rationality assumption of agents, and they overlook the selfish or malicious nature of human users. 
To deal with these issues and challenges, in the proposed modelling approach, we propose a model for MCS which is inherently \textit{dynamic}. 
Indeed, we evaluate the \textit{evolutionary dynamics of human sensing behaviours} through the rounds of \textit{iterated social dilemmas} following a microscopic strategy update rule based on a Fermi statistical distribution (see subsection \ref{ssec:evolutionary}). Thus, the evolutionary game-theoretic approach does not require players/human users to act rationally, but they only choose a strategy at each round of the game trying to maximise their payoff \cite{di2015quantifying}.
In the next section we will describe the proposed \textit{social game-theoretic model}.
\begin{figure*}[!ht]
\centering
\includegraphics[width=1.0\linewidth, height=0.46\textheight, keepaspectratio]{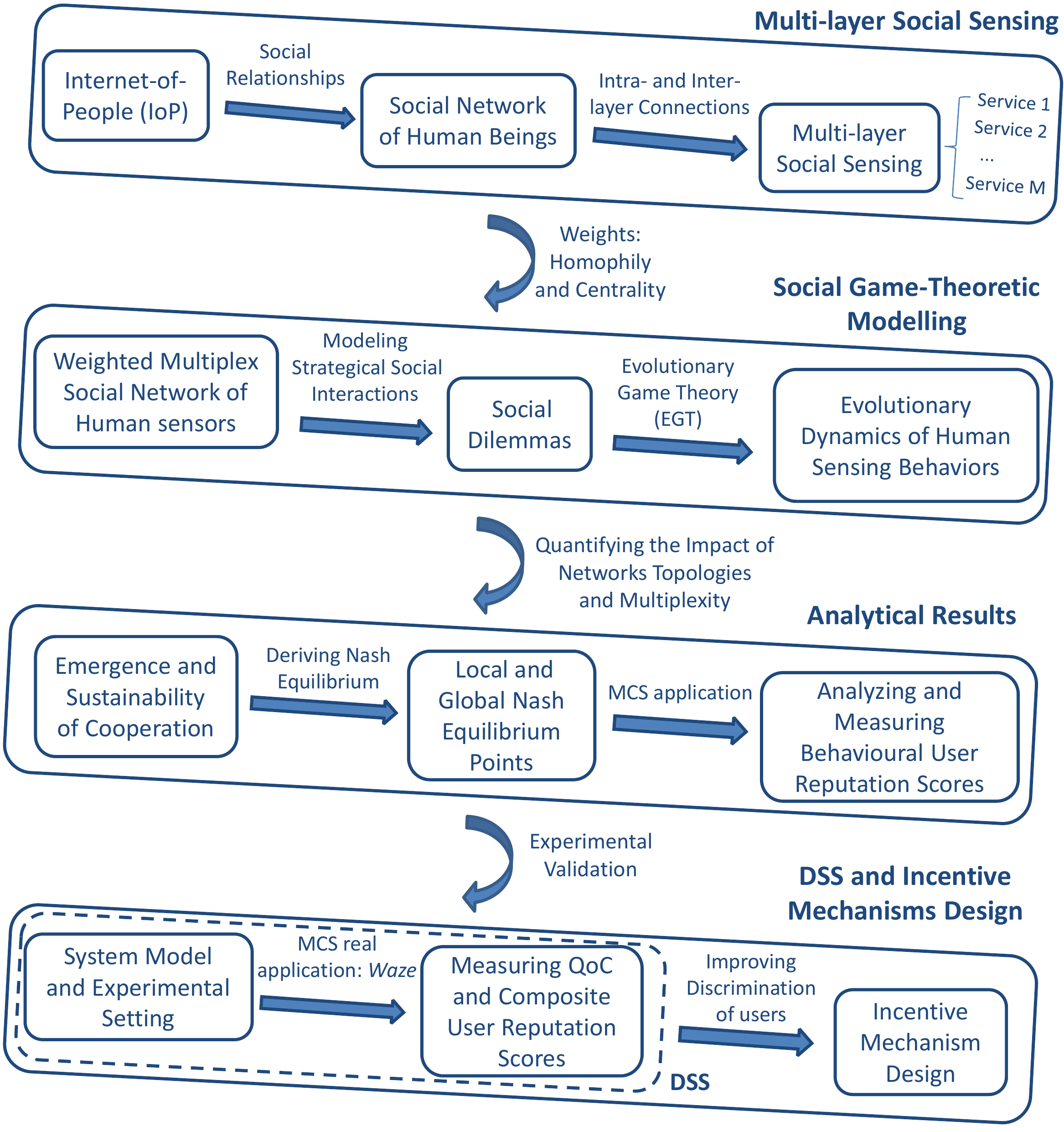}
\caption{{\bf{Social Sensing and Cognitive architecture}.} The figure schematically describes the various steps and aspects of the modelling procedure. Starting from IoP and the definition of the multi-layer social sensing framework of services and weighted social multiplex network of users (first block), we quantify the evolutionary dynamics of human sensing cooperation through game-theoretic modelling (second block). After quantifying the emergence and sustainability of cooperation on various network topologies and by varying homophily between nodes on the network, we derive local and global Nash equilibrium points. Then, based on our analytical model, we analyse and measure the aggregated behavioural user reputation scores for each user in the various configurations (third block). Finally, we conduct the experimental validation on Waze application and we define the composite user reputation scores, the DSS and the design of incentive mechanisms (fourth block) (see Text).}
\label{fig:scheme}
\end{figure*}

\vspace{-0.11in}
\section{Social Game Theory}
\label{sec:model}
The modelling approach is schematically described in Fig. \ref{fig:scheme}. In the first block, starting from IoP we derive the multi-layer social sensing platform, where layers represent the various services but, at the same time, we define also the weighted multiplex social sensing between users, where weighted relationships are those among users and layers are the various channels of social interaction. The second block is the game-theoretic modelling where, starting from weighted social multiplex network, we explore social interactions between users and model the evolutionary dynamics of human sensing behaviours. We quantify the emergence and sustainability of cooperation by varying the network topologies, homophily and multiplexity. Once detected the games and network topologies leading to the emergence and sustainability of cooperation, we analytically define and measure the QoI and the users' behavioural reputation scores in the network deriving from the analytical model. Thus, based on these statistical estimators, we conduct the experimental validation on a real MCS application, \textit{Waze} dataset, by defining and quantifying \textit{Quality of Contribution} (QoC) and \textit{composite user reputation scores}. These measures allow us to define the DSS and incentive mechanism design, measuring its performance.
In this section we present the system model (see subsection \ref{ssec:system}) and then we describe the first two blocks of the framework. The third block is addressed in section \ref{sec:results}, while the fourth block, related to experimental validation, DSS and incentive mechanism design, is described respectively in subsections \ref{ssec:experimental_setting_implementation}, \ref{ssec:DSS} and \ref{ssec:performance}.
\subsection{System Model}
\label{ssec:system}
The system model is a typical urban sensing scenario, where a particular urban area is divided into multiple sensing regions. In each of these regions, there is a population of users along with their smartphones registered to a vehicular crowdsensing application, sending reports or alerts about specific events, such as roadblocks, incidents, road closure, etc. The main components are therefore human users $i$, reports or contributions $j$, and spatio-temporal windows $k$.
As a representative application fitting our discussed system model, we have considered the real-world dataset \textit{Waze} as a case study for the experimental validation of our methodology (for more details about the Waze dataset and its features, please refer to the subsection \ref{ssec:waze} of the Supplementary material). In terms of event occurrence and temporal bias of event distribution we make assumptions similar to \cite{barnwal2019ps}.
In the Waze dataset, events and reports provided by users are divided into spatio-temporal windows, whose definition is based on variations in time of the day and day of the week. We therefore assume that each day and time of the day defines a specific spatio-temporal window.
Human users along with their sensing behaviour contribute data in the form of reports on the status of the traffic at different sensing regions by using smartphone-based PS applications \cite{sajal2020,barnwal2019publish,barnwal2019ps}. Such sensing task may be either explicit (i.e., done directly by humans) or implicit via the sensors (e.g., sensors equipped to smartphones, wearables, vehicles) the humans own \cite{barnwal2019publish}. The contributions sensed by the humans are analysed and aggregated by the vehicular crowdsensing application to trigger and publish an event. This facilitates better decision making in the physical space. In our model, we resort interactions and the mobile crodwdsensing game to derive dynamic measures of QoI and user reputation scores (see subsection \ref{ssec:reputation_score}).

\subsection{Multi-layer Social Sensing and Homophily}
\label{ssec:multiplex_social_sensing}
\begin{figure}[!t]
\centering
\includegraphics[width=1.0\linewidth, height=0.35\textheight, keepaspectratio]{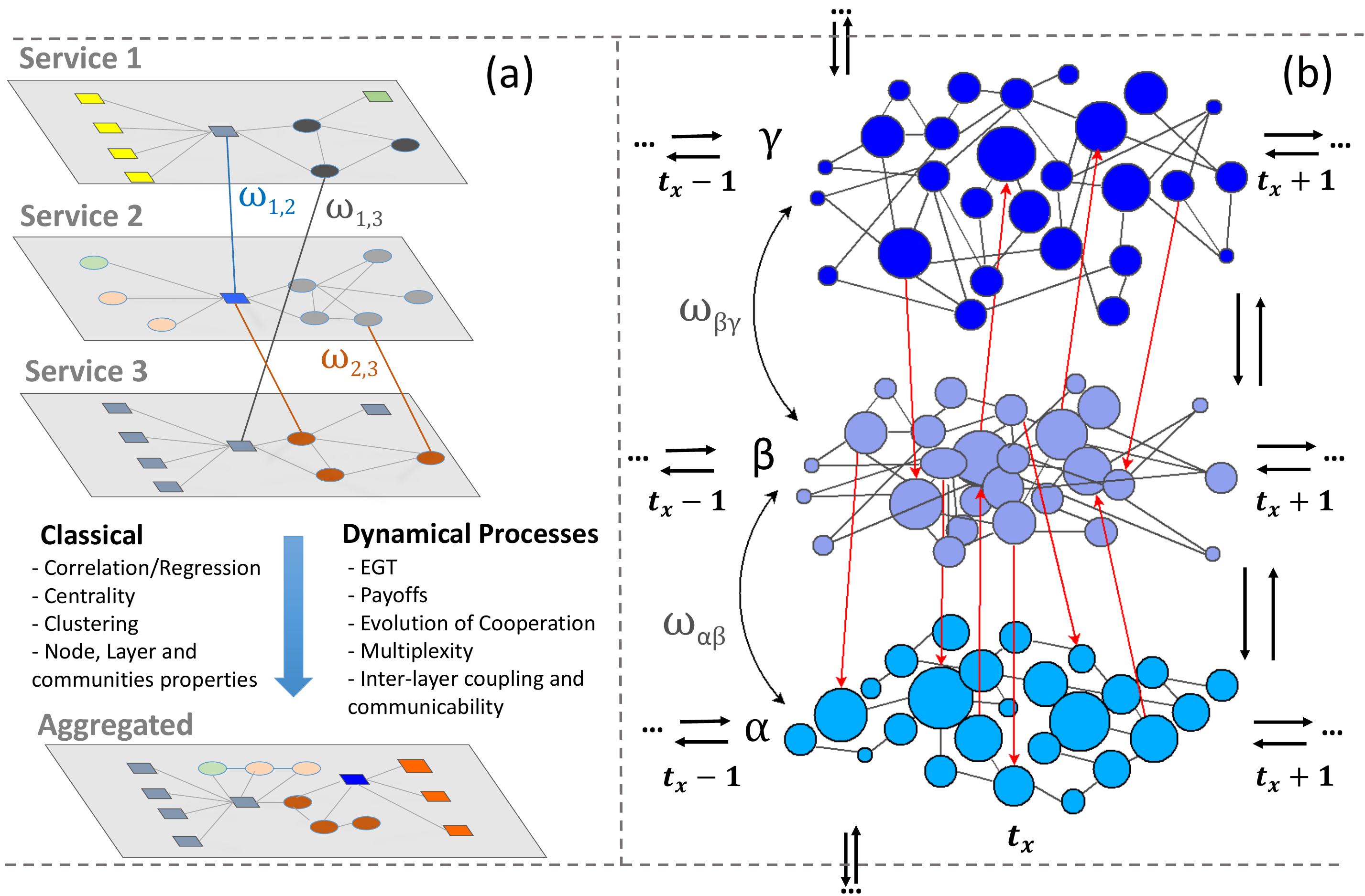}
\caption{{\bf{Multi-layer Social Sensing and inter-layer coupling on a Weighted Multiplex Social Network}.} The figure shows the multi-layer social sensing framework, where layers correspond to different services (see (a)). We show our concept of aggregated network, not only seen classically as including all the aggregated information about node, layer and community properties, but in our model we also include the dynamical aspects, given by evolutionary social dynamics, multiplexity and inter-layer coupling (see (a)). $\omega_{xy}$ is the inter-layer strength between two generic layers $x$ and $y$, respectively corresponding to services (a) and social networks (b) (where layers are denoted by $\alpha$, $\beta$, and $\gamma$). In (b) we illustrate the importance of considering the dynamical patterns of connectivity deriving from the weighted multiplex social network. A perturbation in one layer could drive temporal changes in the other layers through inter-layer coupling between layers in the multiplex structure (see Text).}
\label{fig:multiplex}
\end{figure}

Let us consider a multiplex network of $M$ layers, $\alpha = \{ 1,... , M\}$, and $N$ nodes, $i=\{ 1,... , N\}$, which is a set of $M$ networks ${G_\alpha } = (V,{E_\alpha })$. The set of nodes $V$ is the same for each layer, whereas the set of links $E$ changes according to the layer. Each entity or node in the weighted multiplex social sensing is a human user with a different contribution profile and interacting with other users. 
Fig. \ref{fig:multiplex} describes the multi-layer social sensing modelling approach adopted in this work.

In a MCS environment, layers represent various services exploited by the users, and the multi-layer interactions may affect users' sensing behaviour, and their choice to actively and qualitatively contribute (i.e., cooperate) to the sensing process. Thus, the proposed MCS multi-layer modelling approach is dual: since it is both an IoP-based {\em weighted multiplex social network}, in terms of weighted interactions between users, and a {\em multi-layer social sensing} platform, in terms of services (see Fig. \ref{fig:multiplex}). Indeed, users, while providing and sharing information on various services (see Fig. \ref{fig:multiplex} (a)), interact on the social multiplex network changing their behaviour towards other users in the network and the whole MCS application (see Fig. \ref{fig:multiplex} (b)), thus impacting on the reliability and operation of various services. The aggregated layer conveys all the structural information contained in the multi-layer social sensing.
In Fig. \ref{fig:multiplex} (b) we show the importance of considering the dynamic patterns of connectivity deriving from the coupling between layers of the social multiplex network.
Multiplex networks inherently include the concept of heterogeneity \cite{scata2018quantifying,scata2016impact}, both structurally, in terms of various types of interactions on multiple layers, as well as in terms of contributors' profiles, that is the different behaviours in all the layers of the multiplex structure. 
To characterise and measure such heterogeneity, we define a weight for links between nodes at each layer, so that we consider a weighted multiplex social network, where human behaviours are the result of multi-scale social interactions on such networks \cite{battiston2017new,boccaletti14}.
Weights' definition is based on a combined measure of eigenvector-like centrality and homophily, as defined in \cite{di2015quantifying}. The eigenvector-like centrality measure includes the concept of influence in our analysis. Its definition is based on the spectral properties of the adjacency matrix of each layer of the multiplex network. It allows us to take into account not only the number of links of each node, but also the quality of such connections. Central nodes are the most influential nodes which can influence the behaviours of their neighbouring nodes. 

In our model, other than considering a structural property of network represented by centrality, weights of interactions between human users at each layer of the multiplex structure are defined according to the homophily measure $h_{ij}$. In particular, homophily has a dual definition. On the one hand, it is an interaction-based measure, depending on the frequency of interactions between users on various social channels of interactions in the weighted multiplex social network. Thus, the more they interact on the social multiplex network, the higher will be the homophily measure between them. On the other hand, it is also a similarity-based measure, which takes into account the different dimensions of homophily, along with the actions on MCS services, to derive an Euclidean distance between human users. Overall, by considering both the definitions, we can define the concept of homophily as follows:
\theoremstyle{definition}
\begin{definition}\textit{Homophily}, denoted as $h_{ij}$, is a measure of similarity between two nodes/human users $i$ and $j$, so that:
\begin{equation}
    h_{ij}=\frac{1}{1+\delta_{ij}}
\end{equation}
where $\delta_{ij}$ is the homophily difference (or distance) between users $i$ and $j$ \cite{di2015quantifying}.
\end{definition}

\subsection{Game-Theoretic Modelling - Social Dilemmas and Network Structures}
\label{ssec:dilemmas_topologies}
In order to quantify and capture the social dynamics of human users' behaviours on the social multiplex network, we introduce an evolutionary game-theoretic (EGT) approach. This allows us to obtain a multi-scale analysis of social dynamics and derive the impact of multiplexity on the users' sensing attitude in MCS applications. In particular, we focus on exploring the evolution of cooperation, intended as the emergence and sustainability or resilience of cooperation on the multiplex network. 

In the analysis of human sensing cooperation through EGT, we exploit different social dilemmas where, although these are all two-strategies games, each of them has a different characterisation, reflected by the specific payoff matrix representing its rules of interactions. Hence, social dilemmas allow us to analyse different conflict situations and evaluate how each of them yields distinct Nash equilibrium points and significant changes in game dynamics. 
It is important to observe how a social dilemma is a game which possesses at least one socially inefficient Nash equilibrium. 
In particular, we consider the iterated forms of the Prisoner’s Dilemma game (PD), the Snowdrift Game (SD), the Stag-Hunt game (SH) and the Harmony Game (HG). 
In these social dilemmas, agents/players are the users of a MCS system that, can choose between two strategies: cooperate (C) or defect (D). Cooperating means honestly contributing and participating to the sensing task, such that human user decides to pay a cost of providing his contribution. 
In order to have a high operational reliability in a crowdsensing application and have a robust Quality of Information (QoI), also the other player should contribute. However, there could be users who decide to defect, such as not paying any cost of contribution for accomplishing the task, or relying on the contributions of others in a selfish way. In this case, if also the other player decides to defect, the task will not be accomplished with a negative effect for both players.
A game can be defined in function of its payoff matrix as in Table \ref{tab:payoff}.
\begin{table}[ht] 
\centering
\begin{tabular}{lllll} \cline{1-3} \multicolumn{1}{c}{} & \multicolumn{1}{c}{\bf C} & \multicolumn{1}{c}{\bf D} & & \\ \cline{1-3} \multicolumn{1}{l}{\bf C} & \multicolumn{1}{c}{$R$} & \multicolumn{1}{c}{$S$} & & \\  \multicolumn{1}{l}{\bf D} & \multicolumn{1}{c}{$T$} & \multicolumn{1}{c}{$P$} & & \\ \cline{1-3} & & & & \end{tabular}
\caption{{\bf{Generic payoff matrix}.} Generic payoff matrix of a social dilemma, where $R$, $S$, $T$ and $P$ are respectively the Reward, Sucker, Temptation and Punishment payoffs (see Text).}
\label{tab:payoff} 
\end{table}
Players will both receive a {\em reward} $R$ in the case of mutual cooperation or a {\em punishment} $P$ in the case of mutual defection. A defector will get the {\em temptation} payoff $T$ when playing against a cooperator, while the cooperator obtains the so-called {\em sucker} payoff $S$. 

The difference between the above defined social dilemmas lies in the ranking of payoffs. In the PD, the payoffs are ordered as $T > R > P > S$, meaning that the defection is the best strategy irrespective of the opponent’s decision \cite{matamalas2015strategical}. SD is an anti-coordination game where the payoffs' ranking is the following: $T > R > S > P$, so that it evolves towards the coexistence of both cooperators and defectors. Instead, SH is a classic example of coordination game and the ranking is as follows: $R > T > P > S$. Finally, in the HG the ranking is: $R > S > P$ and $R > T > P$. Overall, the final state of a population playing the HG game will be total cooperation, regardless of the initial fraction of cooperators, the opposite of PD. Moreover, each of the layers of the social multiplex network is a complex network that is represented by a underlying graph where nodes are connected according to a network topology. 
Among them we consider the most investigated network topologies or structures \cite{boccaletti2006complex}: random graphs network models, such as the Erd\"os-R\'enyi (ER) model \cite{erdos1959random}, the Small-World (SW) networks, such as the Watts-Strogatz  model \cite{watts1998collective}, and the Scale-Free (SF) networks \cite{barabasi1999emergence}. We therefore take into account network structures exhibiting a different level of heterogeneity. ER graphs are homogeneous networks, where nodes have the same degree, so that there is a uniform probability to be connected (no degree correlations between nodes in the network) and the degree distribution could be approximated by using the Poisson distribution. SW networks are slightly heterogeneous networks, characterised by a high clustering and modularity, so that there are groups of nodes that are more highly connected than the rest of the network, and there is an over-abundance of hubs (high-degree nodes) that mediate the shortest path length. In SW networks, degree distributions exhibit a fast typically Gaussian decaying tail. Finally, SF networks are highly heterogeneous networks, characterised by a power-law degree distribution. They exhibit a high degree correlation between nodes and degree distribution has a long tail, which means that there are a few hubs in the network.
\vspace{-3pt}
\subsubsection{Evolutionary Dynamics of Human Sensing Behaviours}
\label{ssec:evolutionary}
To explore and quantify the evolutionary dynamics of human sensing behaviours on the social multiplex network, we take into account the iterated forms of the above-described pairwise social dilemmas where, at each round of the game, human users can change their strategies or behaviours, based on imitation dynamics of the fittest strategies \cite{di2015quantifying,scata2016combining}. We quantify and simulate the evolutionary process in accordance with the standard Monte Carlo simulation
procedure, composed of elementary steps, as in \cite{di2015quantifying,scata2016combining,di2019social}, so that at each round a player $i$ changes its strategy $S_{i}$ and adopts the strategy $S_{j}$ from player $j$ with a probability determined by the Fermi function, defined as \cite{di2015quantifying}:
\begin{equation}
\label{eq:fermi}
W(S_{j}\rightarrow S_{i})=(\eta_{i})\cdot \frac{1}{1+exp[(P_{i}-P_{j})/(\delta_{ij}\cdot K)]}
\end{equation}
Therefore, a player $i$ adopts the strategy $S_{j}$ of another player $j$ in function of the payoff difference $P_{i} - P_{j}$, and according to $\delta_{ij}$ and $\eta_{i}$ values. Here $\delta_{ij}$ is the homophily difference between two human users; if this value is small, player $i$ is more likely to imitate the strategy of $j$ at each round. $K$ is the selection intensity and quantifies the uncertainty in the strategy adoption process and it is defined as in \cite{di2015quantifying}. $\eta_{i}$ is the scaling factor defined according to the communicability function between layers of the multiplex structure \cite{di2015quantifying,estrada2014communicability}. It is introduced to include the dependency of the strategy adopted by the player $i$ on the strategies adopted by the counterpart nodes and its neighbours on the other layers \cite{di2015quantifying}. Thus, $\eta_{i}$ points out the coupling between layers $x$ and $y$ ($\omega_{xy})$ in the investigation of the evolution of human sensing behaviours on the multiplex structure and it may result in a bias regarding the strategy adoption of the player $i$ in the subsequent round of the game. Specifically, the scaling factor $\eta_{i}$, is defined as follows:
\begin{equation}
\eta_i = 1 - (\eta_{i_{max}} - \eta_{i_{min}}) \frac{\sum_{j \in\beta,S_j=S_i} [G_{\alpha\beta}]_{ij}}{\sum_{j \in \beta} [G_{\alpha \beta}]_{ij}},
\end{equation}
where at the numerator there is the sum of the communicability functions calculated between the node $i$ on the layer $\alpha$ and all its neighbouring nodes $j$ on the layer $\beta$, adopting the same strategy as player $i$. The denominator represents the sum of the communicability functions calculated between the node $i$ on the layer $\alpha$ and all its neighbouring nodes $j$ on the layer $\beta$ (see subsection \ref{ssec:comm_homophily} of the Supplementary Material for a detailed explanation of the communicability function).

\subsection{Statistical Measures for Designing Incentive Mechanisms - QoI and Behavioural User Reputation Score}
\label{ssec:reputation_score}
In order to design a novel incentive mechanism, we need to define some statistical estimators related to users' behaviour. Thus, starting from the quantification of human users' behaviour, we introduce a novel measure of an aggregated \textit{behavioural reputation score} based on human users' behaviours in the multi-layer social sensing.
In order to define the aggregated behavioural reputation score for each human user, first we quantify the \textit{`social honesty'} of each individual, by defining the statistical estimator $\gamma_{i}$ as follows:
\begin{equation}
\label{eq:gamma}
\gamma_{i}=\sum_{i}(NC)_{i}/(N_{r}*N_{nb}),
\end{equation}
where $NC_{i}$ is the number of cooperative behaviours of node $i$ over the $N_r$ rounds of the game. $N_{nb}$ is the number of neighbours for each node $i$. Thus, $\gamma_{i}$ quantifies the level of cooperativeness of each user in the network, considering its behaviour against neighbourhood and it allows us to classify the contributors in honest, selfish or malicious users \cite{sajal2020,di2019improving}. Here $\gamma_{i}$ ranges in $[0,1]$ such that $\gamma_{i}=0$ reflects a lack of cooperativeness, while $\gamma_{i}=1$ means that a human user has been fully cooperative with a proactive attitude towards its social community and neighbourhood. 
By averaging this measure over the population, we get an overall measure of $QoI$ in the network structure, defined as follows:
\begin{equation}
\label{eq:QoI}
QoI = \gamma_{i}^{av}=(1/N)*\sum_{i} \gamma_{i}
\end{equation}
The aggregated measure of behavioural reputation score $R_{i}$ for each user $i$ is given by the ratio between the local measure of $\gamma_{i}$ for each node averaged over the global attitude of users in the network, given by the QoI measure, so that it is defined as follows:
\begin{equation}
\label{eq:reputation}
R_{i} = \gamma_{i} / \gamma_{i}^{av} = \gamma_{i} / QoI
\end{equation}

The statistical estimator $R_{i}$ quantifies and relates the importance of contribution given by the user as compared with the overall QoI in the network.
Table \ref{tab:estimators} sums up all the statistical estimators and parameters defined in our model.

\begin{table*}[ht] 
\centering
\begin{tabular}{llll} 
\cline{1-2} \multicolumn{1}{c}{\bf Estimator} & \multicolumn{1}{c}{\bf Physical meaning/Definition} & \\
\cline{1-2} \multicolumn{1}{l}{$h_{ij}$} & \multicolumn{1}{c}{Homophily between nodes $i$ and $j$} & \\ 
\cline{1-2} \multicolumn{1}{l}{$\delta_{ij}$} & \multicolumn{1}{c}{Homophily difference (distance) between nodes $i$ and $j$} & \\ 
\cline{1-2} \multicolumn{1}{l}{$\eta_{i}$} & \multicolumn{1}{c}{Scaling factor depending on the communicability function} & \\  
\cline{1-2} \multicolumn{1}{l}{$P_{i}$} & \multicolumn{1}{c}{Payoff obtained by the player $i$} & \\  
\cline{1-2} \multicolumn{1}{l}{$P_{j}$} & \multicolumn{1}{c}{Payoff obtained by the player $j$} & \\  
\cline{1-2} \multicolumn{1}{l}{$S_{i}$} & \multicolumn{1}{c}{Strategy chosen by the player $i$} & \\  
\cline{1-2} \multicolumn{1}{l}{$S_{j}$} & \multicolumn{1}{c}{Strategy chosen by the player $j$} & \\  
\cline{1-2} \multicolumn{1}{l}{$\gamma_{i}$} & \multicolumn{1}{c}{Social honesty of the player $i$} & \\ 
\cline{1-2} \multicolumn{1}{l}{$QoI$} & \multicolumn{1}{c}{Quality of Information} & \\
\cline{1-2} \multicolumn{1}{l}{$R_{i}$} & \multicolumn{1}{c}{Behavioural Reputation Score for each user $i$} & \\ 
\cline{1-2} & & 
\end{tabular}
\caption{{\bf{Statistical estimators of the model}.} We include the definition and physical meaning of the different statistical estimators/parameters of the model (see Text).}
\label{tab:estimators} 
\end{table*}

\section{Analytical Results}
\label{sec:results}
In our model, we aim at evaluating and quantifying the role of homophily, network heterogeneity and multiplexity in the emergence and sustainability of cooperation on the social multiplex network of human users. Thus, the first target is to derive the density of cooperators at steady state in the network.
To this aim, simulations have been conducted choosing a social multiplex network with $M$ layers (where the number of layers has been varied) and $N=200$ nodes or human users, where each layer is modelled as one of the network structures.
Homophily values are randomly chosen following a normal distribution around a mean value, with standard deviation $\sigma$. The target has been to analyse the joint effect of all these aspects in the evolutionary dynamics, detecting those configurations of network structures and social dilemmas more able to make the cooperation among human users propagate and sustain in the network.
\subsection{Density of Cooperators}
\begin{figure*}
\centering
\begin{subfigure}{\textwidth}
\includegraphics[width=\textwidth]{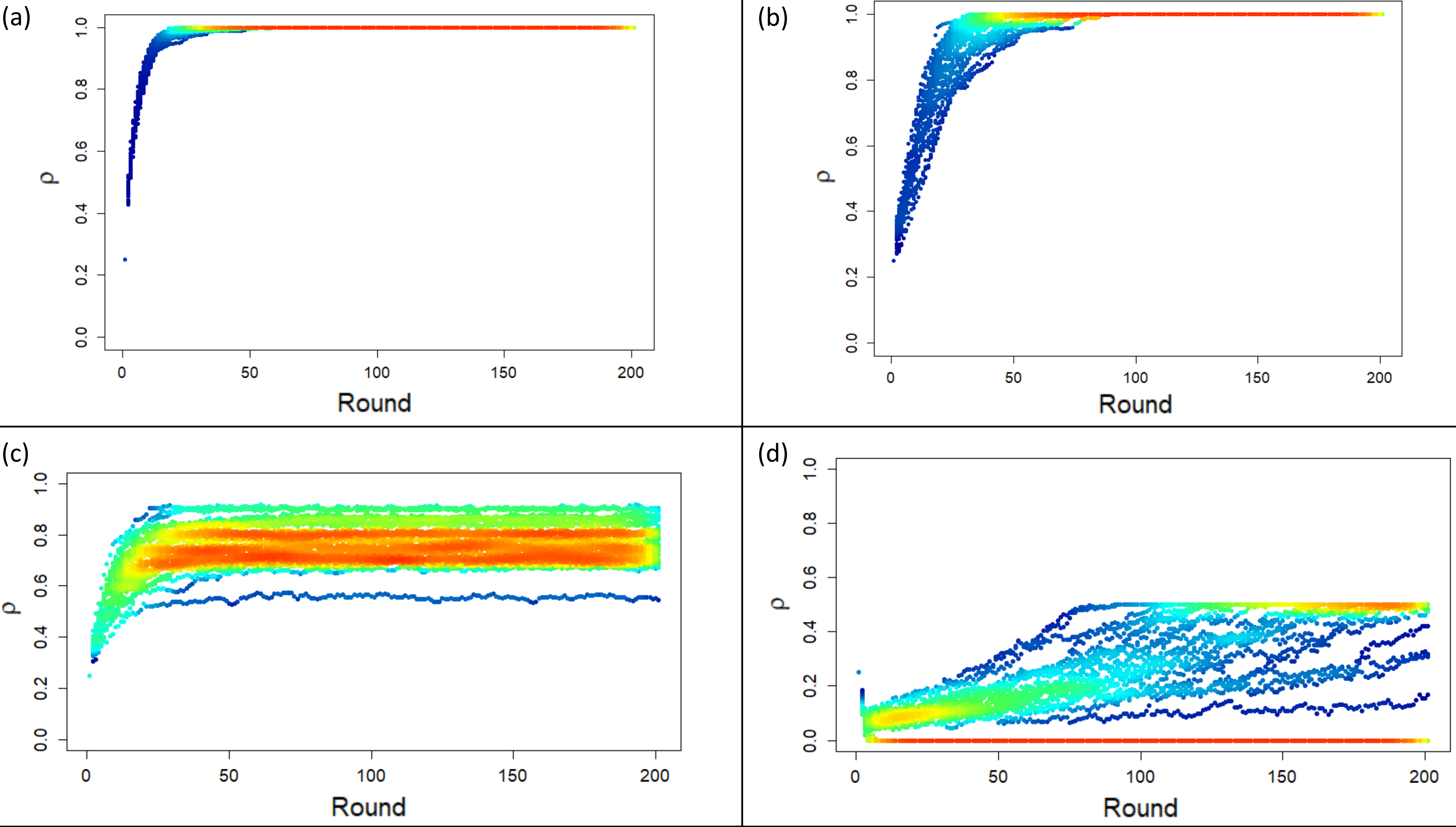}
\end{subfigure}
\begin{subfigure}{\textwidth}
\includegraphics[width=\textwidth]{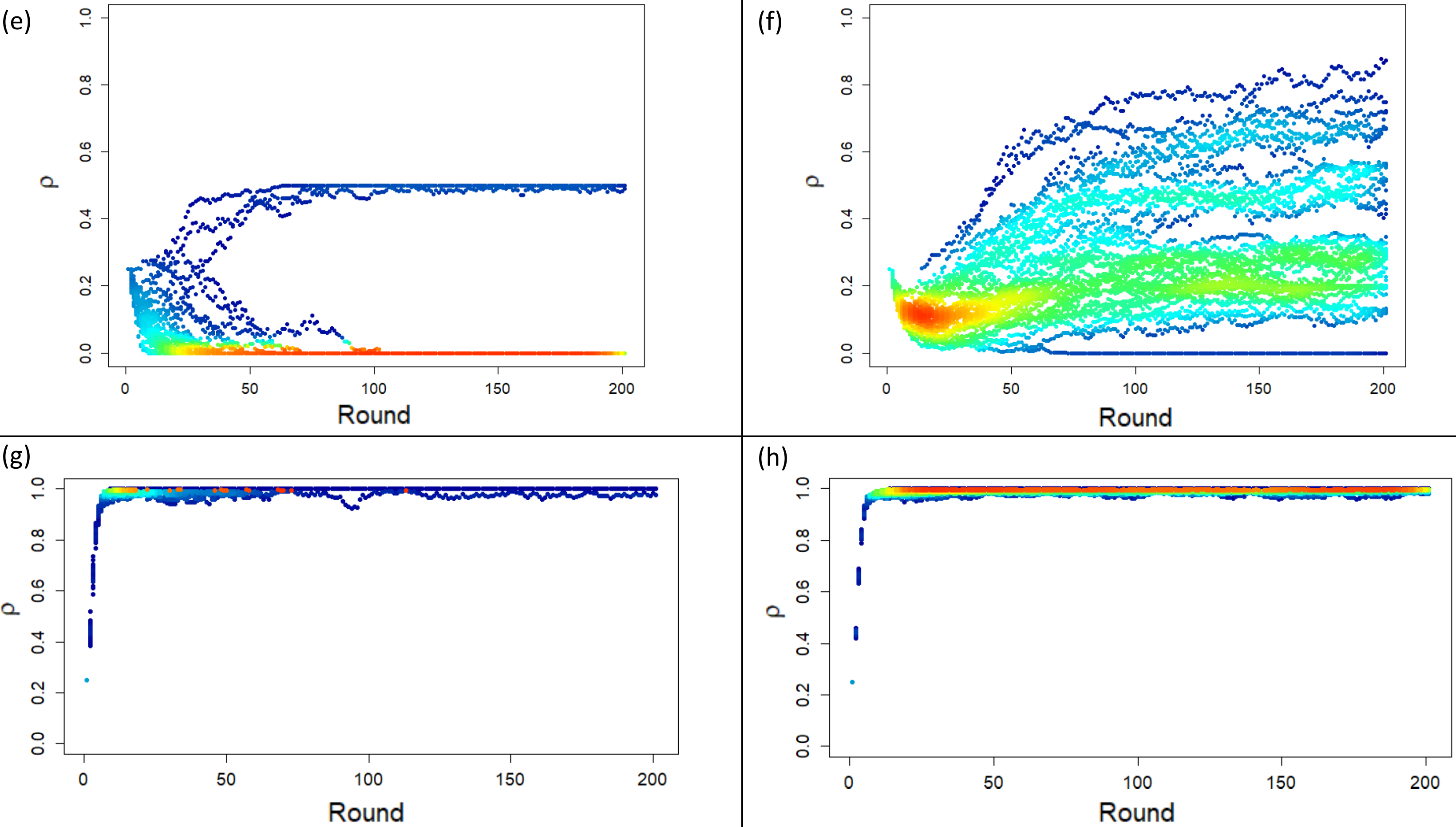}
\end{subfigure}
\begin{subfigure}{\textwidth}
\centering
\includegraphics[width=0.3\textwidth]{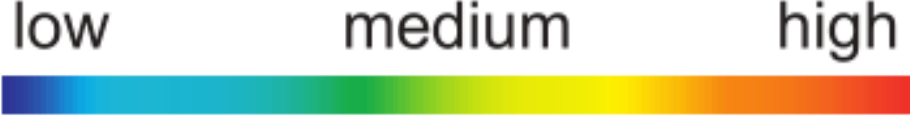}
\end{subfigure}
\caption{{\bf{Density plots}.} The figure illustrates the density of cooperation in various network configurations and social dilemmas (a-h). The colour corresponds to the density $\rho$, ranging from `blue' (lowest) to `red' (highest). We show the following configurations of network structure and game: (a) SF-HG; (b) SF-PD; (c) SF-SH; (d) SW-SH; (e) SW-PD; (f) ER-PD; (g) ER-SD with $M=2$ layers; (h) ER-SD with $M=7$ layers.}
\label{fig:density}
\end{figure*}
Fig. \ref{fig:density} shows the evolution of cooperation, namely the density of cooperative nodes against the rounds of the game. We have simulated the evolutionary dynamics in all the possible configurations of social dilemmas and network topologies, also varying the number of layers, for a number of rounds such that a dynamical steady state was reached.
We can observe how the SF is the most suitable network topology for the emergence of cooperation (see (a)-(b)-(c)). Instead, both in the ER and SW networks, there is a mixed equilibrium with a coexistence of both strategies. More specifically, in the SW network topology, in both cases of SH and PD, we can see the coexistence of both strategies with a prevalence of defectors at equilibrium (see (d)-(e)). The emergence of defection is more marked in the PD rather than in SH, as expected considering the defecting nature of PD. 
While in the ER network configuration, when PD is played between human users, at the beginning we can observe the prevalence of defectors and a coexistence of the two strategies at equilibrium (see (f)). Finally, the last two plots related to ER network and SD (see (g)-(h)) allows us to highlight the role of the number of layers in the evolutionary dynamics of behaviours. In particular, as in \cite{matamalas2015strategical,gomez2012evolution}, an increase in the number of layers results in a stronger emergence of cooperation on the social multiplex network, as indicated by the higher density of `red' points at equilibrium.

This result is what we expected by reasoning in terms of connectivity patterns of the different structures and it is related to the nature of the SF network topology compared with ER and SW networks.
Indeed, SF is inherently heterogeneous, strictly resembling real-world networks displaying a skewed statistical distribution deriving from the preferential attachment rule (\textit{``rich-get-richer''}) \cite{barabasi1999emergence}.
The evolutionary dynamics observed in the case of SW structure derives from its high clustering coefficient of a node with its neighbours, thus bordering the cooperation within communities. 

Comparing the two SF cases (a) and (b), we observe the difference related to the game played by users, where HG produces a higher density of cooperation than in the PD, as expected from literature \cite{matamalas2015strategical}. Thus, in the SF case, although in all the social dilemmas we can notice the emergence of cooperation, HG and SD are the most cooperative dilemmas at evolutionary equilibrium and those more able to sustain cooperation over time. This is even more marked in the high homophily case ($\sigma=1$), where we note a faster emergence of cooperation, rather than in the low homophily case ($\sigma=8$), as expected from \cite{di2015quantifying,scata2016combining}. 

\subsection{Colour Maps}
\begin{figure*}
    \centering
       \includegraphics[width=\linewidth]{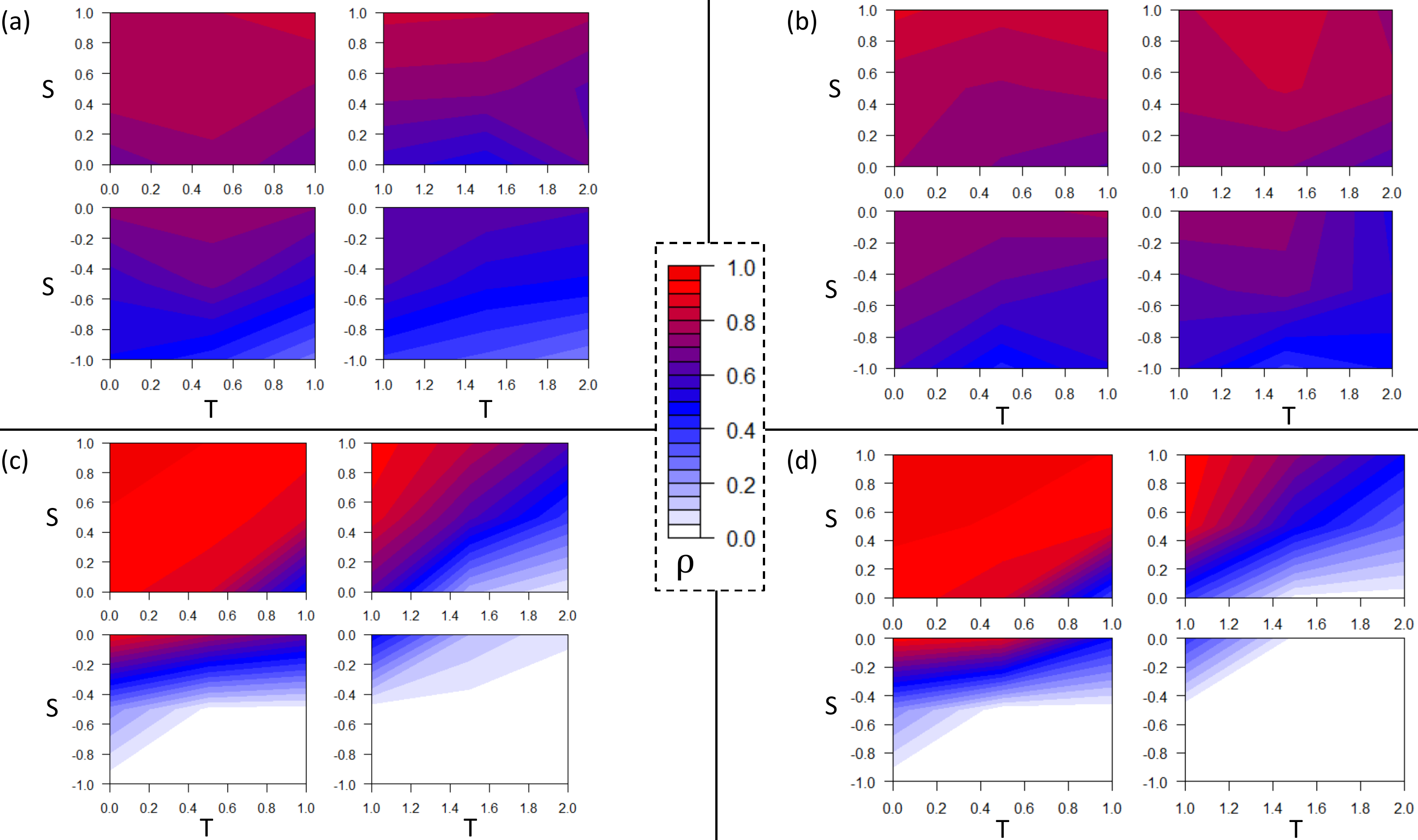}
    \caption{{\bf{Colour maps}.} The figure shows the density of cooperators in the T-S plane, which is divided into four quadrants representing the various social dilemmas: HG (upper-left), (upper-right), SH (lower-left), and PD (lower-right). Plots show the different cases: (a) SF (low homophily), (b) SF (high homophily), (c) ER, (d) SW. Colour corresponds to the density $\rho$, ranging from `white' (lowest) to `red' (highest) (see colour bar).}
   \label{fig:contour}
\end{figure*}
By digging deep into the impact of games along with the various network topologies, in Fig. \ref{fig:contour} we show the density of cooperators in the T-S plane, where each of the quadrants correspond to a different social dilemma. 
Generally, T-S plane allows us to show and better understand what happens in terms of evolutionary dynamics by varying the values of temptation and sucker's payoffs of the different social dilemmas. Furthermore, it allows us to visualise and bring out the transitions areas between the different density levels of cooperation for each network structure and game.
In (a) and (b) we show the most cooperative topology, namely the SF, in both cases of low and high homophily. Despite the different dynamics observed for the different games, in the high homophily case there is a major density of cooperators rather than the low homophily case. Below we discuss the results for each quadrant of the colour maps. 

The upper-left quadrant (HG), where cooperation is its dominant strategy, results the most cooperative game, since we see the higher emergence and resilience of cooperation compared with the other dilemmas, even when choosing high values of temptation payoff ($b$).
The upper-right quadrant (SD) is an anti-coordination game characterised by a stable equilibrium in mixed populations. Thus, we observe the coexistence of both strategies at equilibrium, also varying the values of temptation and sucker's payoffs. In this game the role played by homophily is more evident, since by increasing homophily we see a higher density of cooperation in the T-S plane. In the lower-left quadrant (SH), there is an unstable evolutionary equilibrium with mixed populations. As in the SD case, we have the coexistence of both strategies, even if the density of cooperation is on average lower than SD. Analogously to SD, we can see the increase of density of cooperative nodes according to homophily values. In the PD (lower-right quadrant), defection dominates cooperation, but by comparing the PD in the two cases of low and high homophily, we can see how homophily allows cooperation to emerge. This is more clear for low values of $b$ (say, $b=1$), which is equal to the temptation payoff in the PD, and for values of $c$ that tend towards zero. In the case of PD, an increase of temptation ($b$) and a decrease of sucker's payoff values (since, by increasing $c$ values, we have lower sucker's payoffs, which is equal to $-c$, which means a major cost of cooperating), yield an increasingly lower density of cooperators (see the lower-left corner of the PD quadrant). In cases (c) and (d) we show the density of cooperators in the ER and SW networks. By looking at the density of cooperation in the T-S plane resulting from both these network structures, the previously discussed inherent properties of each game are even more marked than in SF network. Indeed, on one hand we can see the high cooperativeness of HG and vice versa the high density of defecting users in the lower-right quadrant (PD). On the other hand, we can observe how in both SD and SH, we have the coexistence of mixed strategies. 
\subsection{The Role of Network Heterogeneity}
\label{ssec:network_heterogeneity}
Our results shed light on the importance of network heterogeneity, as it induces cooperative agents (or nodes) to create assortative clusters, where they reciprocate cooperation. This mechanism is known as \textit{network reciprocity}, and it represents one of the five mechanisms ruling the cooperation between individuals \cite{nowak2006five}. The main underlying principle is that the benefit produced by cooperating outweighs the cost of cooperating with all neighbours. In particular, by denoting with $r$ the benefit-to-cost ratio (or game return), so that: $r=b/c$, we have that the fixation of cooperation through network reciprocity occurs only if it is satisfied the following condition: $r=b/c > \left \langle k \right \rangle$, where $\left \langle k \right \rangle$ is the average degree in the network (the average number of neighbours). The average degree: $\left \langle k \right \rangle = \frac{1}{N} \sum_{i=1}^{N}(k_{i})$ gives information about the network sparsity, but it does not provide any information on the degree distribution, which is instead the discriminant of each network topology. For this reason, it is crucial to dig deep into the microscopic and mesoscopic issues of cooperation, better explaining how the cooperation evolves \cite{poncela2007robustness}. 
There are mainly three types of players: mixed players (players that change their strategy as the evolutionary process runs), pure defectors and pure cooperators (players that never change their strategy). The organisation of these players is worthy because always exists a boundary of mixed players between pure cooperators and pure defectors. At the mesoscopic level, the community structure of the network typical of many real social networks, has been demonstrated to be important in the preservation of cooperation under heavy temptation to defect conditions.
Reasoning in microscopic terms, as also underlined in \cite{santos2006evolutionary}, the stability of cooperation in the network structure depends on being more in contact with nodes having different and fluctuating strategies and not with the clusters of defectors. The cluster of cooperators is stable if none of its composing nodes has a defector neighbour coupled to more than $N_{C}/b$ cooperators, where $N_{C}$ is the number of cooperators linked to the player. This becomes even more challenging for cooperation to evolve if these connections between a cooperator and a defector neighbour occurs with a high homophily between that nodes.

The main difference among the considered network structures in microscopic terms is related to the distribution and composition of the clusters of cooperators and defectors.
Indeed, in the case of ER graphs, looking at Figs. \ref{fig:density} and \ref{fig:contour} there is a wide region of temptation payoffs with a coexistence of the two strategies, and this is due to the presence of several small clusters of cooperators. This fragmentation of pure cooperators into several clusters of cooperators merged into a region of fluctuating individuals, makes these clusters more exposed to invasion. 
Instead, in SF networks pure cooperators form more compact clusters, that follow hubs behaviours, and finally merging them in a big group or main cluster. The formation over the rounds of games of only one big group of cooperators makes them more resilient to defection as the number of pure cooperators exposed to fluctuating individuals is lower. High homophily values further speed up the process of group (or cluster) formation and also the size of the single cooperative group in the SF network \cite{di2015quantifying}, enhancing the network resilience against an invasion of defectors. In SW networks, there is also a continuous formation of compact clusters of cooperators, due to the high clustering coefficient of SW network. However, these clusters in the network hinder the evolution of cooperation and its fixation in the whole multiplex structure due to the lack of weak ties between groups or cluster of cooperators. Thus, cooperative behaviours remain bounded within clusters, since cooperators in the clusters are less connected to the surrounding network, reducing the probability for a widespread propagation. This is due to the inherent nature of SW networks, having only a small heterogeneity and not highly connected hubs. 
Moreover, it is important to note how the results in terms of evolutionary dynamics of behaviours for the different games are coherent with those obtained in \cite{matamalas2015strategical,gomez2012evolution,di2015quantifying}. In addition, we have analysed and quantified the impact of different network topologies (SF, ER, SW) and the role of homophily, shedding light on its impact on the evolution of cooperation in the various social dilemmas. Results confirm how homophily acts as a shaping factor of cooperation, able to increase the density of cooperation in all the evolutionary games \cite{di2015quantifying,scata2016combining}.

\subsection{Local and Global Nash Equilibrium Points - Homophily and Heterogeneity}
Starting from the previous considerations, related to the various games, network heterogeneity and homophily, in this section, following \cite{zhang2014local}, we define \textit{local} and \textit{global} Nash equilibrium points in our model.
Nash equilibrium represents a key concept in game theory, as it suggests the possible outcomes when different players play simultaneously in order to maximise their payoff. The idea behind equilibrium is that if the players choose strategies that are best responses to each other, then no player has an incentive to deviate to an alternative strategy \cite{easley2010networks}.
In evolutionary settings (EGT), the concept of Nash equilibrium consists of the evolutionarily stable strategy (ESS). Thus, the equilibrium is a stable distribution of strategies, namely a genetically-determined strategy that tends to persist once it is prevalent in a population \cite{easley2010networks}. 

In the case of structured population, such as a multiplex social network, interactions are spatially constrained, as nodes and their counterparts interact only with their neighbours at various layers of the multiplex structure. At each round a node/player plays with only one neighbour as described in section \ref{sec:model}. In our model, links are relatively stable in the whole evolution process, but we change homophily values in order to quantify its impact on different network topologies and games. As showed in density plots (see Fig. \ref{fig:density}) and colour maps (see Fig. \ref{fig:contour}), we observe how the stable states are only three, respectively corresponding to a prevalence of defectors, cooperators and mixed strategies. These stable states are the result of a chain of local Nash equilibrium points \cite{zhang2014local}, until reaching a global equilibrium. Indeed, the overall population is divided into a few groups playing the same strategy, and the various factors, such as network heterogeneity and specifically homophily, impact on the formation of these groups over the time, both in terms of speed and size \cite{di2015quantifying}, as underlined in the previous sections. The higher are the homophily values, the quicker is the formation and size of groups of cooperators, in all the games and specifically their role is more evident in PD and SD games. In other words, we can observe the presence of attractors and a polarisation of strategies mainly focused on defectors and cooperators. Homophily plays a key role in making cooperation percolate through the network, and this is even more marked in a SF network, due to its heterogeneity in the degree distribution as explained in section \ref{ssec:network_heterogeneity}.

Since deriving the exact expressions of ESS or Nash equilibrium points is extremely difficult due to the difficulty of formulating the replicator dynamics, following \cite{zhang2014local}, we define local Nash equilibrium points.
Then, in order to get the ESS, a chain of local Nash equilibrium points is suffice to lead the system into the stable state, since in structured population and heterogeneous networks balancing the gap of payoffs between different strategies is not so difficult.

We are going to discuss and define the local Nash equilibrium in structured populations, exploiting the model presented in \cite{zhang2014local}, and extending their definition including homophily and therefore replacing the adjacency matrix $A_{ij}$ with the matrix $Z_{L}$ as in \eqref{eq:Z_L}. 

At each round of the game, a node/player $i$, plays with its neighbours whose number is quantified by the degree centrality $k_{i}$.
In a two-strategy evolutionary game, we can define the strategy of the player $i$ as follows:
\begin{equation}
\label{eq:theta}
    \Theta_{i}(n) = \binom{S_{i}(n)}{1-S_{i}(n)}
\end{equation}

\noindent
where $S_{i}(n)$ can only take values $1$ or $0$ at the n-th round. For $S_{i}(n)=1$, the player $i$ is a cooperator (C), while for $S_{i}(n)=1$, player $i$ is a defector (D).
Locally, for each player $i$'s gaming environment, we define the local frequency of cooperators at the $n_{th}$ round as follows:

\begin{equation}
\label{eq:Xi}
    \Xi_{i}(n) = \frac{\sum_{j} Z_{ij} \Theta_{j}^{T}(n)\binom{1}{0}}{k_{i}}
\end{equation}

In this scenario, keeping the strategies of the neighbours of the player $i$ unchanged is equivalent to keeping $\Xi_{i}(n)$ unchanged. For the global gaming environment, we define the global frequency (or density) of cooperators at the $n_{th}$ round as follows:
\begin{equation}
\label{eq:rho_n}
    \rho (n) = \frac{\sum_{i} \Theta_{i}^{T}(n)\binom{1}{0}}{N}
\end{equation}

\noindent where $N$ denotes the population, i.e.,  the number of nodes of the social multiplex network.
In a two-player game, the payoff table is a $2 \times 2$ matrix as in Table \ref{tab:payoff}. Considering equation \eqref{eq:theta}, the payoff of player $i$ at the $n_{th}$ round is given by:

\begin{equation}
\label{eq:payoff_i}
    P_{i} (n) =  \Theta_{i}^{T}(n)\binom{R \quad S}{T \quad P} \sum_{j} Z_{ij} \Theta_{j}(n)
\end{equation}

Given eq. \eqref{eq:payoff_i}, $\sum_{j} Z_{ij} \Theta_j(n)$ can be rewritten as:

\begin{equation}
\label{eq:sum_Z_theta}
\sum_{j} Z_{ij} \Theta_{j}(n) = k_i \left( \binom{1}{0} \Xi_{i}(n) + \binom{0}{1} (1 - \Xi_{i}(n)) \right)
\end{equation}

By including the eq. \eqref{eq:theta} and eq. \eqref{eq:sum_Z_theta} into eq. \eqref{eq:payoff_i}, we obtain:

\begin{equation}
\label{eq:Phi}
\Phi_{i}(n) = k_i (\Delta_i(n) S_{i}(n) + T \Xi_{i}(n) +P (1- \Xi_{i}(n)))
\end{equation}

\noindent
where $k_{i}$ denotes the connectivity of node $i$ and $\Delta_i(n)= S-P + (R-T+P-S) \Xi_{i}(n)$. The maximum of $\Phi_i(n)$ is obtained by considering the best strategy, which is denoted by \cite{zhang2014local}:
\begin{equation}
\label{eq:S_i_max}
S_{i,max}(n) = \begin{cases} 1 & , for \quad \Delta_i(n)>0 \\ 1 \quad or \quad 0 & , for \quad \Delta_i(n)=0 \\ 0 & , for \quad \Delta_i(n)<0 
\end{cases}
\end{equation}
If two connected nodes $i$ and $j$ choose strategies $S_{i,max}(n)$ and $S_{j,max}(n)$ as
their strategies at the $n_{th}$ round, respectively, they are in a local Nash Equilibrium. These two nodes/players are called as ``Nash pair''. If all the nodes in the multiplex structure are in the \textit{local Nash Equilibrium}, the population is in a \textit{global Nash Equilibrium}. If $\Delta_i(n)=0$ or $\Delta_j(n) = 0$, this local equilibrium represents a weak local Nash equilibrium. Otherwise, it is a strict local Nash equilibrium. 
The maximum evaluated in the previous equation depends on the value of $\Delta_i(n)$, that is defined starting from the payoff matrix values. In our model, along with payoffs, homophily values impact on the strategy adoption through the $Z_{ij}$ matrix.
Moreover, local Nash equilibrium points represent a series of multiple equilibrium points, whose emergence is ruled by the various factors defined in the model. Each of these local Nash equilibrium points behaves as an agent, and then reaching a sort of coalescence corresponding to the global Nash equilibrium. 

Starting from \cite{zhang2014local}, we redefine a \textit{`Nash pair'}, indicted by $\alpha$, which measures the fraction (or density) of Nash pairs in the overall multiplex network, as follows:

\begin{equation}
\label{eq:Nash_pair}
\alpha =  \frac{N_{p}}{E} 
\end{equation}

where $N_p$ denotes the number of the Nash pairs in a network and $E$ denotes the number of edges through the layers of the network. It is important to note that if two nodes have multiple links on the various layers of the multiplex network, in the aggregated network we only consider this connection once. If $k_i = k_j = 1$, the local Nash equilibrium is equal to Nash equilibrium in the classical game theory. As underlined in \cite{zhang2014local}, the parameter $\alpha$ constitutes a tool to evaluate the evolutionary dynamics of behaviours in the multiplex structure is in an evolutionary stable state.

Local Nash equilibrium points in the structured population are metastable points where sub-populations and groups move towards a stable global equilibrium state over time. This ESS corresponds to the overall cooperation, and its emergence is based on network heterogeneity, the type of game considered and homophily values. The system shows a low multiplicity, a few number of steady states and variously the different factors allow to reach these states. Even though we do not have data to derive the exact trajectory towards the global equilibrium, we observe both analytically and by simulations how an highly heterogeneous network, such as SF \cite{santos2005scale}, allow the cooperation to emerge in the social multiplex network. Furthermore,
high homophily values lead more quickly to local Nash equilibrium points and then the global Nash equilibrium, by speeding up the formation of groups of cooperators, increasing their size and enhancing the network resilience against an invasion of defectors \cite{di2015quantifying}.

\begin{figure*}
    \centering
       \includegraphics[width=0.6\linewidth]{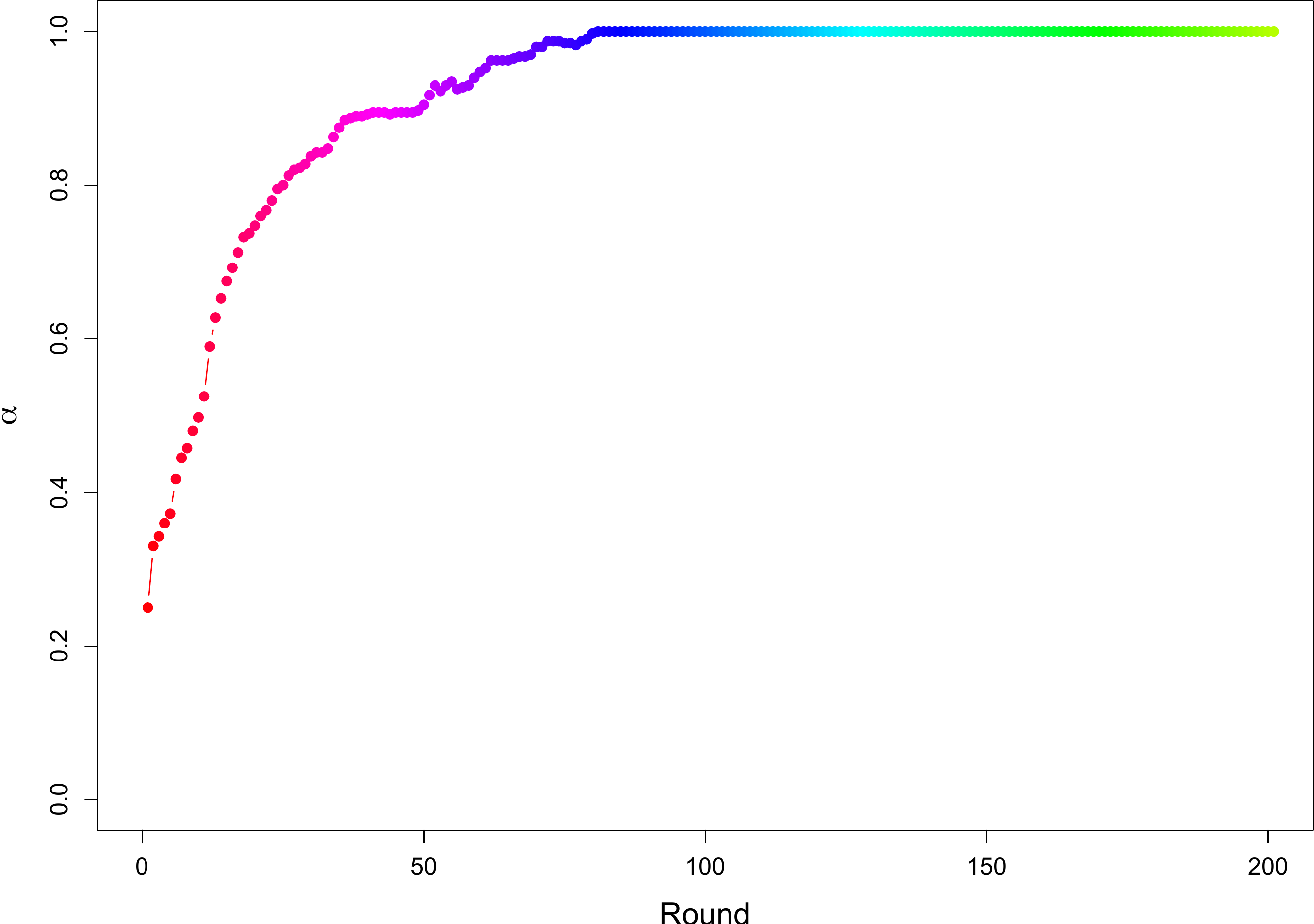}
    \caption{{\bf{Local and global Nash equilibrium points}.} In the case of SF network and PD game, different phase transitions represent the local Nash equilibrium points and finally the global Nash equilibrium represented by the overall cooperation in the multiplex structure (see Text).}
   \label{fig:Nash}
\end{figure*}

Fig. \ref{fig:Nash} shows how the density $\alpha$ of Nash pairs varies over the time, considering high homophily values and a SF as network substrate. We can observe phase transitions with the emergence of local metastable Nash equilibrium points, and finally the global Nash equilibrium given by the overall cooperation.

\subsection{User Reputation Score in a MCS scenario}
\begin{figure*}
   \centering
   \includegraphics[width=\linewidth]{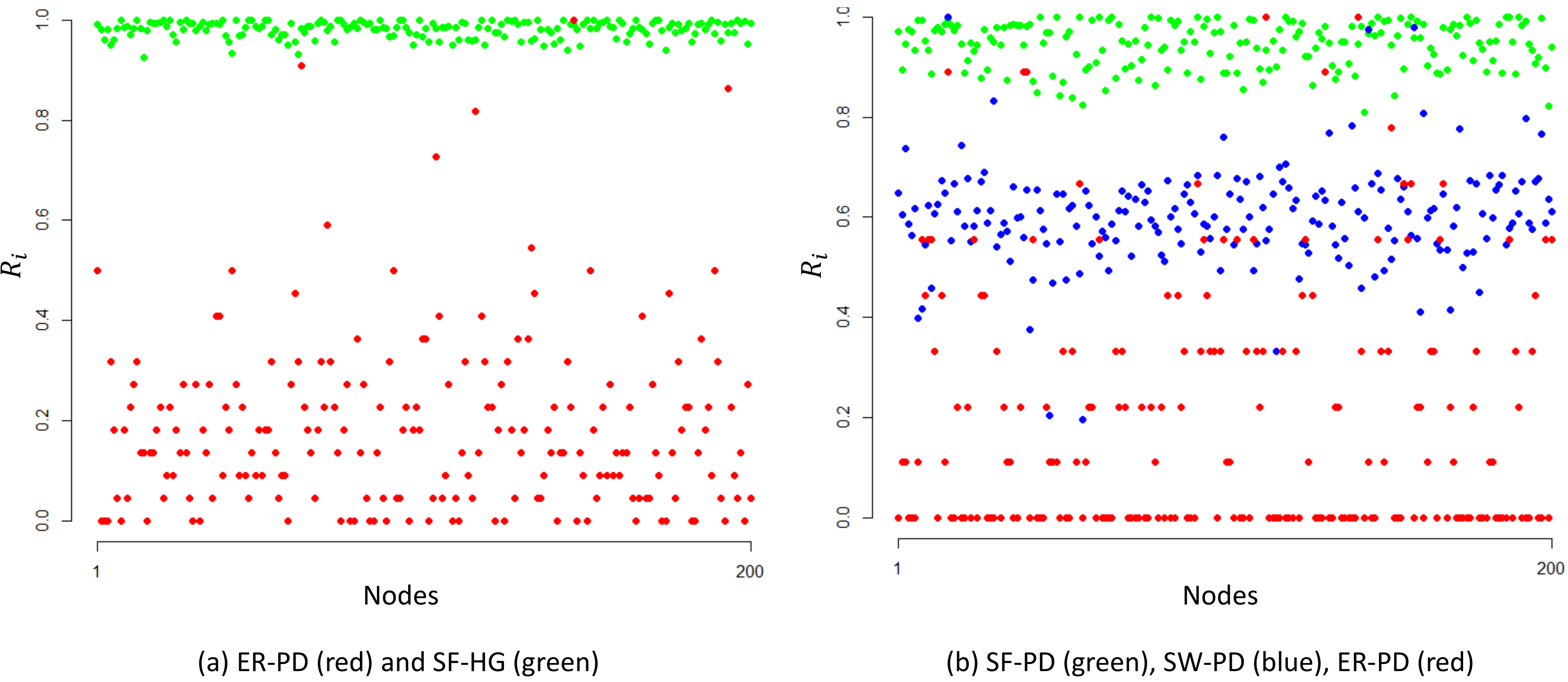}
    \caption{{\bf{Behavioural User Reputation scores}.} The figure shows the user aggregated behavioural reputation scores (y-axis) against nodes (x-axis) (a) in two configurations of network structure and social dilemmas, and (b) a comparison between the behavioural user reputation scores deriving from different network structures considering the same social dilemma (PD) (see Text).}
   \label{fig:user_reputation}
    \end{figure*}

We have simulated a simple scenario of vehicular crowdsensing applications (services) using synthetic generated following an approach similar to that in \cite{sajal2020}. We have considered synthetic data of a simple scenario of a vehicular crowdsensing application, but our modelling approach is generic and may be extended to any kind of MCS service. We formulate the MCS in vehicular networks and the interactions between human users/vehicles equipped with sensors as a vehicular crowdsensing game. Each of them participates to the sensing task and chooses its strategy based on some constraints related to sensing costs/risks and gains derived from the accomplishment of the sensing task.
As discussed earlier, we consider both the quantity (degree of participation) and quality (accuracy of contributions) of the reported events \cite{sajal2020,di2019improving}. They represent the contributors' profiles and have been used as input data. Then, following the proposed game-theoretic modelling approach, we have measured the behavioural user reputation scores for each human node in the different configurations of network structure and social dilemmas. 
By looking at Fig. \ref{fig:user_reputation}, we can observe how in the SF case, behavioural user reputation scores assume high values distributed between $0.85$ and $1.0$, significantly higher than in the ER and SW cases. Indeed, in the ER and SW cases scores are mainly distributed in the range between $0$ and $0.4$. We can also observe how the ER network behaves worse than SW in terms of behavioural user reputation scores.

\section{Experimental Validation}
\label{sec:experimental_validation}
Starting from the system model defined in subsection \ref{ssec:system}, we present its implementation and experimental settings (see subsection \ref{ssec:experimental_setting_implementation}), mapping the statistical estimators of our model onto the data extracted from the Waze dataset. Then, we describe the DSS and the incentive mechanism design (see subsection \ref{ssec:DSS}), and we derive the experimental results related to incentives' disbursement (see subsection \ref{ssec:performance}).

\subsection{Experimental Setting and System Model Implementation}
\label{ssec:experimental_setting_implementation}
Based on our social game-theoretic model, we extract from the Waze dataset also the definition of \textit{neighbours}, and we assume that neighbours are those users indirectly interacting in the same spatio-temporal window thorough the Waze application. In other words, two users interact with each other if they report an event in the same spatio-temporal window.
Furthermore, in order to distinguish the quality of reports, and derive a measure of truthfulness of human user contribution, we use the attribute \textit{Report Rating}, and we assume that a report is of high quality (cooperation) if \textit{Report Rating} is strictly greater than the average measure of Report Rating.
We define the truthfulness $\tau_{ij}^{k}$ for any contribution $j$ at time $k$, by modelling the problem as a weighted regression approach, where the predictor variables are: (i) the expectation from the opinion about a contribution and (ii) the spatial distance between the GPS-based location of the user $i$'s smartphone and the cellular tower-based location. Truthfulness is a dependent variable deriving from these two terms. From our Waze dataset, the measure of the truthfulness of a report corresponds to the \textit{Report Rating}.

\textbf{Quality of Contribution.}
Once defined $\tau_{ij}^{k}$ representing the expected truthfulness on a report $j$, contributed by user $i$ at time $k$, now, we need to define the \textit{Quality of Contribution} (QoC).
As in \cite{barnwal2019publish}, we map the truthfulness values $\tau_{ij}^{k}$ to a QoC value through the generalised linear models (GLM). Thus, by applying the regression through the logit function on $\tau_{ij}^{k}$, we have:
\begin{equation}
\label{eq:QoC}
    Q_{ij}^{k}=ln \left( \frac{\tau_{ij}^{k}}{1-\tau_{ij}^{k}} \right)
\end{equation}
$Q_{ij}^{k}$ is defined in the range $[-\infty, + \infty]$ and it allows us to determine whether the odds of the contribution $j$ is true or false. The logit function is a monotonically decreasing function giving lower weights to $\tau_{ij}^{k} < 0.5$.

In our model, we extend the concept of QoC, which is related to the truthfulness of a contribution, by including a human user behavioural measure represented by the `social honesty' $\gamma_{i}$ of each user (see eq. \eqref{eq:gamma} in section \ref{ssec:reputation_score}). It allows us to redefine the QoC as follows:
\begin{equation}
\label{eq:QoC_gamma}
    \overline{Q_{ij}^{k}}= \gamma \cdot ln \left( \frac{\tau_{ij}^{k}}{1-\tau_{ij}^{k}} \right)
\end{equation}
where $\overline{Q_{ij}^{k}}$ is obtained by multiplying the previous QoC by the `social honesty' $\gamma_{i}$ of each user. By quantifying the degree of cooperation of user $i$ in the previous rounds, it represents a measure of local `reputation' of the user $i$, that amplifies his QoC. Indeed, in the measure of $\gamma_{i}$, we include the number of cooperations and we weigh this value based on the number of neighbours. Thus, we evaluate not only the quality and quantity of reports generated by user $i$ over the time, but also the local \textit{spatio-temporal distribution or density} of cooperative reports over the overall number of reports at each window. In other words, we weigh the importance of a report in a certain window based on the relative percentage of high-quality reports generated in that specific spatio-temporal window. This gives us a measure of spatio-temporal density of cooperation, and the importance of a contribution is inversely proportional to the density of cooperative reports: the lower is the density, the most critical is that spatio-temporal window, and the higher is the incentive to be assigned to that user.
Moreover, $\gamma_{i}$ entails a measure of \textit{persistence} of cooperation, namely the number of times or spatio-temporal windows where a user contributes a high-quality report (i.e., Report Rating strictly greater than the average of Report rating).

\textbf{User Reputation.}
In Section \ref{ssec:reputation_score}, we have derived the measure of Quality of Information (QoI) by averaging $\gamma_i$ over all the $N$ users in the network (see eq. \eqref{eq:QoI}), and an aggregated behavioural reputation score $R_{i}$ deriving from $\gamma_{i}$ for each user $i$.
Now, we define the \textit{composite user reputation score} $RS_{i}$ by including both the expected truthfulness of report $j$ generated by user $i$ at time $k$, that is $\tau_{ij}^{k}$, and the newly defined measure of QoC derived also from users' degree of cooperation in the complex network (see eq. \eqref{eq:QoC_gamma}).
By aggregating the QoCs of the contributions generated by each users in the network, we have the composite user reputation score:
\begin{equation}
\label{eq:overall_RS}
    RS_{i} = \sum_{k=1}^{T} \overline{Q_{ij}^{k}} \cdot p_{ij}^{k}
\end{equation}
\noindent where $p_{ij}^{k}$ is equal to $1$ if the user $i$ has generated a contribution (or report) $j$ at time $k$, otherwise it is equal to $0$. Given that the aggregated measure of the composite reputation score $RS_{i}$ can assume any value in the interval $[- \infty, + \infty]$, we normalise it in order to have values in the interval $[0,1]$.

\subsection{Decision Support System (DSS) and Incentive Mechanism Design}
\label{ssec:DSS}
\begin{figure}[!t]
\centering
\includegraphics[width=1.0\linewidth, height=0.5\textheight, keepaspectratio]{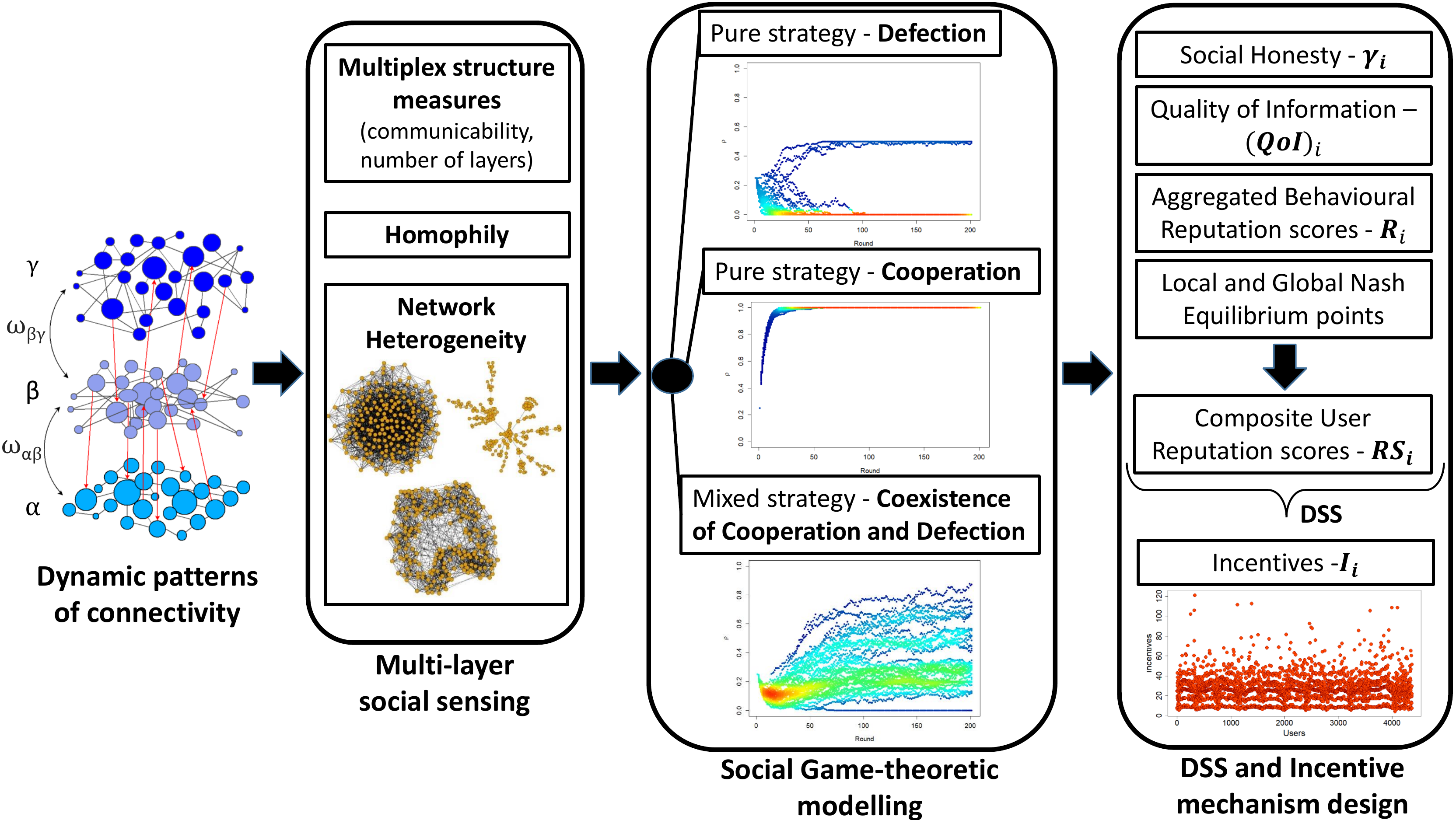}
\caption{{\bf{Social game-theoretic modelling for designing incentive mechanisms in MCS}.} Conceptual description of the proposed methodology, which has been also implemented in the software. Starting from the dynamic patterns of connectivity on a multi-layer social sensing framework, which includes homophily, network heterogeneity (network structures in the block correspond to ER, SF and SW networks), the social game-theoretic modelling leads us to quantify the truthfulness measures, and define the DSS and the incentive mechanism design.}
\label{fig:final_scheme}
\end{figure}

In Fig. \ref{fig:final_scheme} we conceptually map the various aspects of the proposed methodology onto the MCS space. Starting from the dynamic patterns of connectivity, the multi-layer social sensing framework, which includes homophily, network heterogeneity and multiplex structure measures, and the game-theoretic modelling guides choices according to human sensing behaviours. It leads to evaluate and quantify the human-centric policies, defining a MCS space that allows us to quantify the truthfulness measures for designing incentive mechanisms. 
We therefore get a Decision Support System (DSS) able to perform a decision making process related to disbursing incentives to users based on dynamic and human-centric policies. As in \cite{muller2019personalisable}, these policies represent a minimised set of rules extracted from both qualitative and quantitative information and data related to human users and their behaviours. Thus, the DSS results from the analysis and quantification through a social game-theoretic modelling approach which allows us to join and combine multiplexity, network heterogeneity and human-related factors represented by homophily.

Here we describe more in detail the DSS and its decision making model on a specific event based on contributions (or reports) provided by human users in a vehicular crowdsensing scenario, starting from the measures defined in the previous section. 
Then, based on our novel definition of user reputation, we define the design of incentive mechanisms.

\textbf{Decision Support System.}
As described in Section \ref{ssec:system}, the system is able to distinguish honest from selfish and malicious users by computing their QoC and user reputation, where the defined measures of QoC (eq. \eqref{eq:QoC_gamma}) and user reputation (eq. \eqref{eq:overall_RS}) encompass the evolution of user behaviours in the network. Thus, both measures are inherently dynamic and derive from the evolutionary dynamics of user sensing behaviours in the network.
Now, we show how our proposed methodology can be applied to infer true events and support decisions for publication of events.
Although the definition of user reputation derives also from the truthfulness of user contributions, the decision making process results challenging in the absence of the ground-truth. The three main concerns are that different types of event reports that can be received within in the same sensing region, which may be relevant or irrelevant in reality. This is mainly due to the human perception difference. In addition, the lack of previous interactions, with new users (without any reputation score) may make the issue even more problematic. Finally, some users can build higher reputation first, and then start orchestrating fake events.
Mistakes in decision making may lead to publishing fake events or dropping true events and it may result in the wasteful expenditure on incentives (losses) to the malicious users. Furthermore, these mistakes also decrease the operational reliability of the overall crowdsensing application.

To avoid these wrong decisions, the DSS must be able to decide at runtime which event among multiple reported events is most likely to be accurate. Inspired by  \cite{sajal2020,barnwal2019publish}, we define a two-level DSS, where we present a simpler and lightweight model based on \cite{sajal2020}. The first level is linked to decide \textit{what} type of event to publish, that is evaluating the most likely event type that has occurred. This first decision level is required due to the presence of contributions or reports received simultaneously in the same sensing region but related to more than one event type. As in \cite{barnwal2019publish}, a confidence value for each reported event type $j$ is computed based on the relative quantity and quality support for each event type $j$.
The second level must determine \textit{whether} to publish or drop an event. To this aim, there should exist sufficient evidence to suggest that the most likely event has actually occurred. Based on the extent of evidence, this decision level therefore allows us to discriminate if the publication of this event will result in a benefit or gain. This aspect aims to prevent orchestrated fake events, since honest reporters will not report anything in the absence of any event.
Based on the Waze dataset, we assume that among the reported events, the occurred event corresponds to the mostly reported event by only considering high-quality reports.

Let the Waze application receive reports from different users in a common spatio-temporal window. We know the users' reputation scores from the last spatio-temporal window, and for a specific event the MCS application aggregates all these reports, that we denote by $N_{agg}(j)$. Let us denote by $RS_{agg}(j)$ the aggregated reputation score of all users reporting the $j$-th event type. Let us denote with $U^{+}(z)$ the total number of users with a positive reputation score ($RS_{i} \geq 0.5$) currently present in the spatio-temporal window. Mathematically, we model the overall \textit{confidence} on any event type $j$ ($C_{j}$) using a weighted sum of quantity and quality supporting each event type $j$.

\begin{equation}
     C_{j} = \nu \cdot \frac{N_{agg}(j)}{U^{+}(z)}
     + (1-\nu) \cdot \frac{RS_{agg}(j)}{\sum_{i \in U^{+}(z)} RS_{agg}(j)}
\end{equation}

Therefore, for $j$-th event, the first term represents the relative support for quantity, since it quantifies the relative participation to the event $j$, while the second term represents the quality.
The weight $0 \leq \nu \leq 1$ is the \textit{preference factor} associated with the evidence types, and it is tuned based on available contextual spatio-temporal information of a given type of event or risk policy.
In our model, the main difference with \cite{sajal2020} is that the reputation score $RS_{i}$ derives not only from the quality and quantity of contributions provided, but also including behavioural and topological or network structural aspects. Indeed, we calculate user reputation scores (see eq. \eqref{eq:reputation} and eq. \eqref{eq:overall_RS}) from the evolutionary dynamics of sensing behaviours in the multi-layer social sensing framework, which includes homophily and network structural heterogeneity (see section \ref{sec:model}).

\textbf{Incentive Mechanism.} Based on the above sections, let us describe our incentive mechanism in mobile crowdsensing applications which is inherently dynamic and fair. Specifically, incentives received by user $i$ are given by:

\begin{equation}
\label{eq:incentives}
    I_{i} = \frac{RS_{i}}{\sum_{m=1}^{U^{+}} RS_{m}} \frac{B \cdot U^{+}}{U}
\end{equation}

where $RS_{i}$ is the aggregated reputation score of user $i$, $U^{+}$ is the number of users in the system with positive reputation score (namely, $RS_{i} \geq 0.5$), $B$ is the total budget allocated for incentives, and $U$ is the total number of users or the population in the system. The ratio between the reputation score of user $i$ and the reputation scores of all the other users showing a positive reputation score (on average behaving mostly as cooperators in the network) represents the relative reputation of the user $i$ in the overall system. It acts as a discounting factor to the maximum possible incentive that any user in the network can gain. Users with higher relative reputation scores will end up getting higher rewards.
Table \ref{tab:measures} sums up all the measures  defined in the system model, DSS and incentives mechanism design and used in the experimental validation.

\begin{table*}[ht!] 
\centering
\begin{tabular}{llll} 
\cline{1-2} \multicolumn{1}{c}{\bf Measure} & \multicolumn{1}{c}{\bf Physical meaning/Definition} & \\
\cline{1-2}
\\
\multicolumn{1}{l}{$\overline{\tau_{ij}^{k}}$} & \multicolumn{1}{c}{Truthfulness of report generated by user $i$ on event $j$ at time $k$} & \\
\cline{1-2} 
\\
\multicolumn{1}{l}{$\overline{Q_{ij}^{k}}$} & \multicolumn{1}{c}{Quality of Contribution of user $i$ on event $j$ at time $k$} & \\ 
\cline{1-2} \multicolumn{1}{l}{$RS_{i}$} & \multicolumn{1}{c}{Composite Reputation Score of user $i$} & \\ 
\cline{1-2} \multicolumn{1}{l}{$C_{j}$} & \multicolumn{1}{c}{Confidence on any event type $j$} & \\  
\cline{1-2} \multicolumn{1}{l}{$I_{i}$} & \multicolumn{1}{c}{Incentive for each user $i$} & \\  
\cline{1-2} & & 
\end{tabular}
\caption{{\bf{Measures of the system model}.} We include the definition and physical meaning of the various measures defined in the system model, DSS and incentive mechanism design.}
\label{tab:measures}
\end{table*}

\subsection{Incentivisation of Users}
\label{ssec:performance}
\begin{figure*}[!t]
    \centering
       \includegraphics[width=\linewidth]{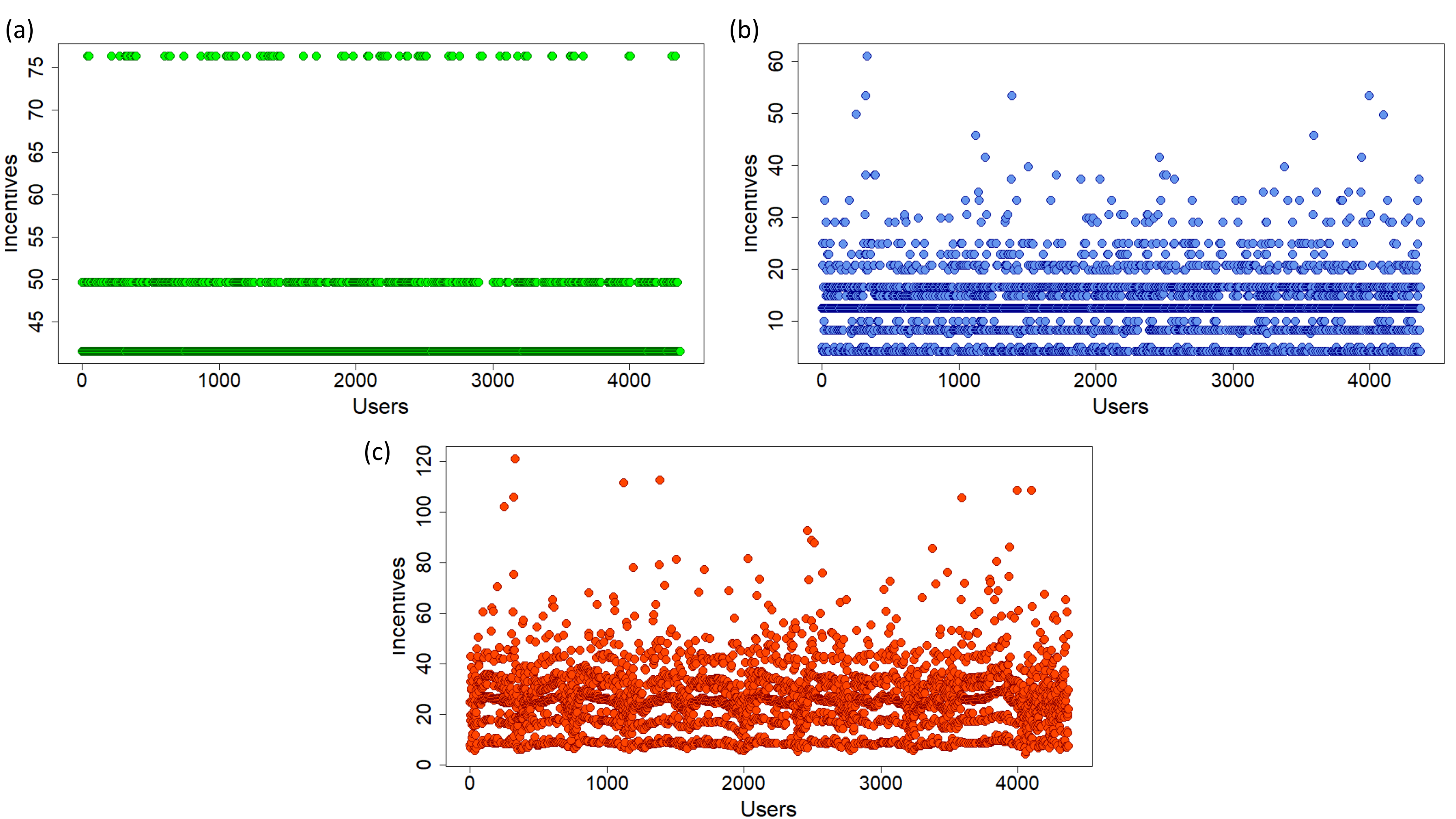}
    \caption{{\bf{Incentives disbursement}.} Disbursement of incentives in the three cases: (a) based on the definition present in \cite{sajal2020,barnwal2019publish}; (b) considering persistence of cooperation; (c) considering both persistence and the local spatio-temporal distribution of cooperators.}
   \label{fig:incentives}
\end{figure*}

In our methodology (see section \ref{sec:model}) we have underlined the importance of including a social game-theoretic approach in designing policies in a MCS scenario. We have included the concepts of homophily, network structural heterogeneity and we have underlined the key role of network and indirect reciprocity for making cooperation emerge and sustain in the network. 
For this reason, in the definition user reputation scores and incentives (see subsection \ref{ssec:DSS}), along with quality and quantity of contribution, we have included the social honesty of users. This has led us to provide a novel definition of reputation scores and incentives that focus also on social dynamics and behavioural aspects derived from human cooperation. Indeed, our definition includes also the number of times a user cooperates in the various windows, or behavioural \textit{persistence}, and the \textit{spatio-temporal distribution (or density) of cooperation} in the different windows. 
These two novel concepts derive from the previously conducted social game-theoretical analysis and all the considerations linked with network and indirect reciprocity \cite{nowak2006five}. Moreover, the introduction of these two novel aspects in the design of incentive mechanisms allow us to translate the idea of indirect reciprocity and detect the `super-cooperative' users \cite{nowak2011supercooperators}. These users cooperate in the most critical windows and the MCS application needs to incentivise them to preserve the operational reliability of service.

In Fig. \ref{fig:incentives}, we show experimental results obtained in terms of disbursement of incentives considering the different incentive mechanisms. Specifically, following the incentive mechanism based on quality and quantity \cite{sajal2020}, but without including neither persistence nor the local spatio-temporal distribution of cooperators (see (a)), we can distinguish only three levels or classes of users' incentives. By including the persistence, we note how incentives are more differentiated, and the classification and distinction of users' incentives is even more marked if we also consider the local spatio-temporal distribution of cooperators (see (c)). The reason is that in (c), along with the quality and quantity of provided reports and a macroscopic measure of persistence of cooperation over time, we have included also local and microscopic aspects. This leads to a higher discrimination of users and a distribution expressing differences between users both at a cooperation and social scale, that is also considering the social groups user belong to.
Thus, the proposed mechanism avoids not only incentives losses deriving from disbursing incentives to selfish or malicious users \cite{sajal2020}, but also those losses linked with a lack of a further discrimination of users which include social network and behavioural factors, such as persistence and the spatio-temporal distribution of cooperative behaviours.

\section{Discussions}
\label{sec:discussions}
In MCS application scenarios, based on participatory sensing, human behaviours assume increasingly a crucial role in the operational reliability of the MCS services. Incentive mechanisms are therefore designed to motivate users to contribute focusing both on the quality and quantity of contributions \cite{sajal2020}. In this work, we have proposed a methodology which includes the definition of the Decision Support System (DSS) and the design of a novel incentive mechanism, extracting the rules and human-centric policies or metrics to disburse incentives to human users. Other than simultaneously participate to various services, they are social nodes that interact each other on a weighted social multiplex network, where each layer corresponds to a distinct type of relation among them and weight of each connection depends on homophily between nodes.


Our model is both data-driven and model-driven. Indeed, starting from data related to different MCS services, we can estimate the statistical measures and define a novel incentive mechanism. On the other hand, from a model perspective, the proposed model defines a class of models, that can be adapted and refined according to specific rules and metrics. Furthermore, by considering human factors and a game-theoretic approach, the proposed methodology aims to make a MCS system increasingly efficient, resilient and adaptive.

Starting from our statistical estimators and methodology, and the proposed novel definitions of the DSS and incentive mechanism, we can get a substantial improvement in terms of policies, as the innovation action may resort these network and game-theoretic aspects, making the entire MCS system more robust and resilient, and ending up stimulating pro-social behaviours of users thanks to reciprocity.

We envisage that the proposed methodology, enclosing social dynamics, multiplexity and human-related issues, may provide new insights in the future design of socially-aware and human-centric incentive mechanisms.
Moreover, our incentive model paves the way and defines the guiding principles for generative models since it provides parameters (statistical estimators) for generating new benchmarks with a fine stratification or classification of users based on local social groups. Indeed, by including macroscopic and microscopic characterisation of users our methodology is able to incorporate both the case of observing different behaviours in similar groups and the opposite case of observing similar behaviours in distinct groups. This leads to a higher level of resolution on different scales and this may become even more marked if we consider a multi-layer network of services, where each layer represents a different service and users participate simultaneously to different services.

\section{Conclusions}
\label{sec:conclusions}
Mobile crowdsensing is a people-centric sensing system based on users' contributions and incentive mechanisms, aiming at stimulating them both in quantity and quality, play a key role for the operational reliability of the system.
In this work, we have re-designed incentive  mechanisms by introducing a social game-theoretic methodology. To this aim, we have defined a multi-layer social sensing framework, where humans acting as social sensors interact on multiple social layers and various services.
We have analysed such interactions and quantified the role of homophily, network heterogeneity and multiple interactions in shaping human sensing behaviours on a networked scenario. We have therefore defined a game-theoretic modelling approach which allows us to get a multi-scale integration of all these issues, quantifying the emergence and resilience of cooperation on the weighted social multiplex network.
In this way, it has been possible to detect which configurations of both social dilemmas and network structure bring out the emergence and sustainability of cooperation.
Then, by considering a simple scenario of MCS and synthetic data related to a vehicular crowdsensing scenario as input data of our model, we have measured the user reputation scores in a synthetic social multiplex network, based on the configurations detected in the previous evolutionary dynamics. Findings have confirmed how user reputation scores are higher for those configurations more able to make cooperation emerge and sustain in the network. 
We have therefore proposed to include also social dynamics, multiplexity and human-related issues in the design of socially-aware and human-centric incentive mechanisms.

Then, starting from Waze, a real-world dataset on vehicular traffic management, we have conducted experiments on the disbursement of incentives by also comparing our method with baselines. We find out how our incentive mechanism is able to better discriminate user behaviours based on both quality and quantity of reports (global) other than the local or microscopic spatio-temporal distribution of behaviours.

In future works, we will include a dynamic definition of weights and payoffs related to micro-packets of energy exchanged between nodes, i.e., micro-affirmations and micro-inequities \cite{di2019social}, to better understand the evolutionary dynamics of the human sensing behaviours. Moreover, we aim at further validating the efficacy of the proposed methodological approach on a multi-layer social sensing, by considering also other real datasets of different MCS applications, in order to analyse the impact of interdependence.
\section*{Acknowledgements}
The authors are grateful to the editor and reviewers for insightful comments that helped improved the technical quality of the manuscript significantly.
This work was partially supported by NSF grants CNS-1818942 and CCF-1725755 and the Research Grant: Italian Ministry of University and Research (MIUR) - PON REC 2014 - 2020 within the project ARS01 01116 ”TALIsMAN”.


\section{Supplementary Material}
\subsection{Communicability Function}
\label{ssec:comm_homophily}
Overall, the factor $\eta_{i}$ enables us to measure the effect of inter-layer coupling and influence between a node/player and its replica on the other layers. Thus, we leverage the definition of communicability function provided in \cite{estrada2014communicability}, which quantifies the number of possible routes that two nodes $i$ and $j$ in the multiplex have to communicate with each other. 
Therefore, considering a multiplex network consisting of $M$ layers, denoted by $L_{1}, L_{2}, ... , L_{M}$, and their respective matrices $Z_{1}, Z_{2}, ... , Z_{M}$, representing the Hadamard product between the homophily matrices and the adjacency matrices of the multiplex $\mathbb{M}$, its matrix is then given by: $\mathfrak{M} = Z_{L} + C_{LL}$, where $Z_{L}$ is defined as follows:
\begin{equation}
\label{eq:Z_L}
Z_L = \oplus_{a = 1}^{M}Z_{\alpha} 
\end{equation}
and $C_{LL}$ is a matrix describing the inter-layer interaction, defined as follows:
\begin{equation}
\label{eq:C_LL}
C_{LL} = \begin{bmatrix}0 & C_{12} & ... & C_{1M} \\
C_{21} & 0 & ... & C_{2M} \\
\vdots  & \vdots  & \ddots  & \vdots \\ 
C_{M1} & C_{M2} & ... & 0\end{bmatrix} \in {R^{NM \times NM}}
\end{equation}
where each element $C_{\alpha\beta} \in \mathbb{R}^{N \times N}$ represents the interaction of layer $\alpha$ with layer $\beta$. Here it is assumed that we have a symmetric interaction between layers, that is: $C_{\alpha\beta}= C_{\beta\alpha} = C = \omega_{\alpha\beta} I = \omega_{\beta\alpha} I$, for all layers $\alpha$ and $\beta$. $\omega$ is the parameter describing the strength of the inter-layer interaction, and $I \in \mathbb{R}^{N \times N}$ is the corresponding identity matrix. So we can now write the multiplex matrix as follows:

\begin{equation}
\label{eq:multiplex}
\mathfrak{M} = \begin{bmatrix}
Z_1&\omega_{12}I & ... & \omega_{1M}I\\
\omega_{21}I & Z_2 & ... & \omega_{2M}I \\
\vdots  & \vdots  & \ddots  & \vdots \\ 
\omega_{M1}I & \omega_{M2}I &... & Z_M
\end{bmatrix} \in \mathbb{R}^{NM \times NM}
\end{equation}

As we are interested in accounting for all the walks between any pair of nodes in the multiplex, we take into account the number of walks of length $k$ between two generic nodes $i$ and $j$ in the multiplex, which is given by the $\alpha$, $\beta$-entry of the $K$-th power of the adjacency matrix of the network. Hence, the walks of $k$ length in the multiplex are given by the different entries of $\mathfrak{M}^{K}$. These walks can include both intra- and inter-layer hops \cite{estrada2014communicability} and we are interested in giving more weight to the shortest walks than to the longer ones. 

The communicability between two nodes $i$ and $j$ in the multiplex network is defined as the weighted sum of all walks from $i$ to $j$ as follows:
\begin{equation}
\label{eq:G_xy}
G_{ij} = I + \mathfrak{M} + \frac{\mathfrak{M}^2}{2!}+... = \sum_{k=0}^\infty \frac{\mathfrak{M}^k}{k!}=\left[exp(Z_L + C_{LL})\right]_{ij}
\end{equation}

We can now define the communicability matrix $G$, where each element $G_{\alpha\beta} \in \mathbb{R}^{N \times N}$ is the matrix representing the communicability between each pair of nodes belonging to two different layers $\alpha$ and $\beta$, of the multiplex $\mathbb{M}$. It is defined as follows:
\begin{equation}
\label{eq:G}
G = exp(Z_L + C_{LL}) = \begin{bmatrix}
G_{11} & G_{12} & ... & G_{1M} \\
G_{21} & G_{22} & ... & G_{2M} \\
\vdots  & \vdots  & \ddots  & \vdots \\ 
G_{M1} & G_{M2} & ... & G_{MM} \\
\end{bmatrix} 
\end{equation}

where $G \in \mathbb{R}^{NM \times NM}$ and $[G_{\alpha\beta}]_{ij}$ represents the communicability between the node $i$ on the layer $\alpha$ and the node $j$ on the layer $\beta$. 

\subsection{Waze Dataset}
\label{ssec:waze}
In the experimental validation, we have considered the real-world dataset available from Waze alerts. It consists of traffic incident alerts generated between 23-February-2015 and 1-March-2015 by the users of Waze application. It has approximately $22,910$ users, $71,505$ reports generated for various traffic incidents occurring in $991$ streets or locations across 11 boroughs of Boston City, Massachusetts (MA). This dataset consists of multiple features, however, for the target of this work, we have considered the following ones:

\begin{itemize}
    \item \textit{Object ID}: ID that uniquely identify a report
    \item \textit{Report Generation Date}: date on which the report is generated
    \item \textit{DayTime}: time (UTC) on which the report has been included in the Waze database
    \item \textit{Street}: name of the street where the traffic incident has occurred
    \item \textit{Incident Type}: type of traffic incidents - accident, jam, weather hazard, and road closure
    \item \textit{UUID}: a 16-octet number generated by Waze to uniquely identify the device from which a given report is generated
    \item \textit{ReportRating}: Rating (between 0 to 5) assigned to each generated report by the Waze application server
\end{itemize}

It is important to observe that to avoid duplicate/redundant reporting, as in \cite{barnwal2019ps}, we have discretised the duration of a day into $8$ segments, each with a span of $3$ hours (early morning, morning, day, mid-day, evening, late evening, night, and midnight).
Moreover, as observed in \cite{barnwal2019ps},
most of the users, except a few outliers, have generated around three reports per week. However, there are a few users who have generated a huge number of reports (almost a report at each spatio-temporal window). These are false or spam reports either due to network outages or malicious behaviour and Waze server has assigned a Report rating equal to zero due to the lack of authenticity. In our pre-processing and cleaning of the dataset we have filtered out these outliers and missing values.
Furthermore, there are also a few users generating duplicated reports in the same spatio-temporal window and we have excluded also these reports from our analysis because they result not significant for our purposes.

For a more detailed description of the dataset and some statistics on frequency of generated reports and distribution of reports in the different days of the week and day times, please refer to \cite{barnwal2019ps}.

\end{document}